\documentclass[12pt,preprint]{aastex}

\usepackage{epsfig,lscape} \usepackage[dvipdfm]{hyperref} \usepackage{natbib}
\newcommand{\myemail}{linlin@mail.ustc.edu.cn, zouhu@nao.cas.cn,
xkong@ustc.edu.cn}

\begin{document}

\title{Gradients of Stellar Population Properties and Evolution Clues in a
Nearby Galaxy M 101}

\author{Lin Lin \altaffilmark{1,2}, Hu Zou \altaffilmark{3}, Xu Kong
\altaffilmark{1,2}, Xuanbin Lin \altaffilmark{1,2}, Yewei Mao
\altaffilmark{1,2}, Fuzhen Cheng \altaffilmark{1,2}, Zhaoji Jiang
\altaffilmark{3}, Xu Zhou \altaffilmark{3}}

\altaffiltext{1}{Center for Astrophysics, University of Science and Technology
of China, Hefei 230026, China; \myemail} \altaffiltext{2}{Key Laboratory for
Research in Galaxies and Cosmology, Chinese Academy of Sciences, Hefei 230026,
China} \altaffiltext{3}{National Astronomical Observatories, Chinese Academy of
Sciences, Beijing 100012, China}


\begin{abstract} 

Multi-band photometric images from ultraviolet and optical to infrared are
collected to derive spatially resolved properties of a nearby Scd type galaxy M
101. With evolutionary stellar population synthesis models, two-dimensional
distributions and radial profiles of age, metallicity, dust attenuation, and
star formation timescale in the form of the Sandage star formation history are
obtained. When fitting with the models, we use the IRX-$A_\mathrm{FUV}$
relation, found to depend on a second parameter of birth rate $b$ (ratio of
present and past-averaged star formation rate), to constrain the dust
attenuation. There are obvious parameter gradients in the disk of M101, which
supports the theory of an ``inside-out" disk growth scenario. Two distinct disc
regions with different gradients of age and color are discovered, similar to
another late-type galaxy NGC 628.  The metallicity gradient of the stellar
content is flatter than that of H {\sc ii} regions. The stellar disk is
optically thicker inside than outside and the global dust attenuation of this
galaxy is lower, compared with galaxies of similar and earlier morphological
type. We highlight that a variational star formation timescale describes the
real star formation history of a galaxy. The timescale increases steadily from
the center to the outskirt. We also confirm that the bulge in this galaxy is a
disk-like pseudobulge, whose evolution is likely to be induced by some secular
processes of the small bar with relatively young age, rich metal, and much
dust.  

\end{abstract}

\keywords{Galaxies: individual(M 101) --- Galaxies: evolution --- Galaxies:
photometry --- Galaxies: bulge}


\section{INTRODUCTION}

Galaxy formation and evolution are complicated matters which have not yet come
to a complete and coherent understanding. In the cosmological $\Lambda$ cold
dark matter ($\Lambda$CDM) model, galaxies are formed by collapsing and cooling of
gas. Galaxies gain in mass and determine their shape and structure through 
clustering and merging. This model has been very successful at reproducing 
observations in the formation of massive galaxies at high redshift \citep{dad05,tru06}. 
In recent years, however, the role of secular evolution on nearby galaxies 
has been found to be more and more significant, compared with the violent 
merging and clustering in the early universe \citep{kor04,car07,fis10,kor10}. 
A lot of observational evidence shows that evolutions of bulge and disk are being 
leaded by some secular processes, which are caused by the asymmetric 
gravitational potential from oval, bar, or spiral structure\citep{kor04}.

For a galaxy, a variety of processes might act during its formation and
evolution, such as merging, satellite accreting, star forming, and radial
migrating \citep{wys06,ros08}. The consequent long-time star formation and
chemical enrichment would lead to non-homogeneous stellar population properties
(e.g., age and metallicity). In general, nearby elliptical and spiral galaxies
are bluer outwards \citep{pel90,tam03}, dwarfs show redder outskirts
\citep{cam09}, and the disks of majority of spiral galaxies have radial abundance
gradients.

Spectral line indices (Lick system) are effective tracers of stellar populations 
for early-type galaxies \citep{wor94}. But for late-type galaxies, 
it is difficult to obtain reliable measurements due to low surface brightness 
and contaminated stellar continua by gaseous emission lines of star-forming 
regions. Thus, many published studies focus on gradients of stellar population 
properties derived by the emission lines. \citet{zar94} and \citet{van98} found 
H {\sc ii} regions to have an average metallicity gradient of -0.05 dex 
kpc$^{-1}$. Recent deep investigations with absorption-line spectra have 
highlighted the presence of non-monotonic colors and stellar population 
gradients in spiral galaxies \citep{mac04,bak08,mac09,san09}. However, gradients 
measured by different authors have large discrepancies, because it is difficult 
to disentangle the effects of age, metallicity, and dust extinction.

Multi-wavelength photometric data can help us to break this kind of degeneracy.
Observed stellar light is emitted by stars in galaxies and might be 
reprocessed by gas and dust in the surrounding interstellar medium. So UV-to-IR 
spectral energy distributions (SEDs) contain a large amount of information about 
galaxies, such as the stellar population, stellar mass to light ratio, and 
behaviors of the gas and dust contents. We have analyzed spatially resolved stellar 
population properties for two nearby galaxies using different evolutionary 
population synthesis (EPS) models \citep{kon00,li04,zou11a,zou11c}. In 
these papers, optical or UV-to-IR SED for each part of galaxies was extracted 
from multi-band photometric images so as to obtain various parameters by fitting 
with models. Our purpose is to investigate the nature and connection of disks 
and bulges, analyze detailed stellar population properties, and get hints on 
the star formation history and galactic evolution.

In this paper, we present a two-dimensional analysis of a nearby galaxy M 101.
M 101 (NGC 5457, $\alpha = 14^\mathrm{h}03^\mathrm{m}12.5^\mathrm{s}, 
\delta = +54^\circ20{\arcmin}56{\arcsec}$) is a face-on Scd 
galaxy at a distance of 7.4 Mpc. It has an apparent scale of about 36 pc per 
arcsec, which provides an excellent opportunity for anatomizing underlying 
stellar populations. The paper is structured as follows.  Section \ref{sec2} 
briefly describes optical photometric observations, archival data in 
other bands, and corresponding data reduction. Section \ref{sec3} explains 
the way to construct model SEDs. In Section \ref{sec4}, we detailedly 
introduce the technique to constrain dust attenuation so as to degrade 
the degeneracy among different parameters. The fitting method and its 
reliability are provided in Section \ref{sec5}. The distributions and 
radial profiles of different parameters are shown in Section \ref{sec6}.  
At last, Sections \ref{sec7} and \ref{sec8} respectively give some
discussions and conclusions.  We use the major-axis position angle of
39$^\circ$ and inclination angle of 18$^\circ$ \citep{bos81} to deproject the
galaxy throughout our paper.

\section{OBSERVATIONS AND DATA REDUCTION \label{sec2}}

\subsection{Optical Intermediate-band Observations} 
M 101, as one of galaxies in the Beijing-Arizona-Taiwan-Connecticut (BATC) survey, 
was observed by the 60/90 cm Schmidt telescope at the Xinglong Station of National
Astronomical Observatories of China (NAOC). A 2048 $\times$ 2048 CCD is mounted
at the focal plane of the telescope. The field of view (FOV) is about
58{\arcmin} $\times$ 58{\arcmin} and the pixel scale of the CCD is about
1.7{\arcsec}. The BATC photometric system includes 15 intermediate-band filters
with bandwidths of about 200$-$300 \AA, covering the wavelength range of
$3300-10000$ \AA. These filters are designed to avoid the contamination 
from night sky emission lines \citep{fan96}. Multiple exposures for each band 
were taken from 1995 November to 2004 February. Table \ref{tab1} lists the 
effective wavelengths of those filters and some statistics of the 
stacked images.

All observed images are treated by the data processing procedures
customized for the BATC survey as described in the paper of \citep{fan96}. The
pipelines include several steps as following: a) overscan correction; b)
subtractions of bias and dark current; c) flat-field correction; d) image
stacking of multiple exposures, where bad pixels and cosmic rays are removed;
e) astrometry by the UCAC3 (USNO CCD Astrograph Catalog) catalogue; f) flux
calibrations of the stacked images by standard stars observed at photometric
nights (for full knowledge of the calibration, refer to the paper of
\citet{zou11c}).  Finally, the CCD digital counts in ADU are converted to flux
density in erg s$^{-1}$cm$^{-2}$Hz$^{-1}$.

\subsection{Archival Data in Other Bands} \subparagraph{GALEX \& XMM-OM
ultraviolet images} GALEX imaging observations are centered at 1529 {\AA} and
2312 {\AA} for the far-ultraviolet (FUV, 1350$-$1750 \AA) and near-ultraviolet
(NUV, 1750$-$2750 \AA) bands, respectively \citep{mar05}. M 101 was observed in
2008 April and May as part of the public Guest Investigator (GI) program, with
exposure times of 13294 s for FUV and 13293 s for NUV. The PSF FWHMs of the
images are $\sim$4.2{\arcsec} and $\sim$4.6{\arcsec}, and sensitivity limits (1
$\sigma$) for these two bands are 2.04 $\times$ 10$^{-19}$ and 1.03 $\times$
10$^{-19}$ erg s$^{-1}$cm$^{-2}$\AA$^{-1}$ per pixel. More details about the
GALEX data are given in \citet{bia05}.

The XMM Optical Monitor (XMM-OM) also observed this galaxy in six broad bands:
VUW1, UVM2, UVW2, $U$, $B$, and $V$ \citep{kun08}. The central wavelengths are
2120 \AA, 2310 \AA, 2910 \AA, 3440 \AA, 4500 \AA, and 5430 \AA, respectively.
Considering the signal-to-noise ratio (S/N), we only use the UVW1 and $U$ band
images to reinforce the observation data in the NUV wavelength region. The
total exposure times of these two bands are 2317 s and 927 s, which yield
1$\sigma$ sensitivity limits of 1.15 $\times$ 10$^{-30}$ and 1.89 $\times$
10$^{-30}$ erg s$^{-1}$cm$^{-2}$Hz$^{-1}$ per pixel. Note that a small part
(west) of M 101 was not covered in the $U$ band.

\subparagraph{2MASS near-infrared images} Near-infrared (NIR) images of M 101
in the $J$, $H$, and $K_s$ bands (1.2, 1.65, and 2.2 $\mu$m, respectively) are
obtained from the 2MASS Large Galaxy Atlas \citep{jar03}.  The FWHMs of the
mosaics are estimated to be about 2.5{\arcsec} and the sensitivity limits
(1$\sigma$) are 3.29$\times$10$^{-29}$, 5.59$\times$10$^{-29}$, and
5.89$\times$10$^{-29}$ erg s$^{-1}$cm$^{-2}$Hz$^{-1}$ per pixel.

\subparagraph{Spitzer infrared images} As part of the Local Volume Legacy
Survey \citep{lee08,ken07}, M 101 was observed in 2004 April by both the
Infrared Array Camera (IRAC; 3.6, 4.5, 5.8, and 8.0 $\mu$m) and the Multi-Band
Imaging Photometer (MIPS; 24, 70, and 160 $\mu$m). Four IRAC mosaics provide a
view of the old stellar population and/or emission from polycyclic aromatic
hydrocarbons (PAHs). The FWHM of each mosaic is about 2{\arcsec} and the
sensitivity limits at 1$\sigma$ are 2.72 $\times$ 10$^{-30}$, 3.47 $\times$
10$^{-30}$, 1.36 $\times$ 10$^{-29}$, and 1.24 $\times$ 10$^{-29}$ erg
s$^{-1}$cm$^{-2}$Hz$^{-1}$ per pixel. Three MIPS bands trace thermal radiations
from the warm and cool dust. The MIPS mosaics have measured FWHMs of about
6.0{\arcsec}, 18{\arcsec}, and 40{\arcsec}. Sensitivity limits at 1$\sigma$ are 2.35
$\times$ 10$^{-29}$, 1.84 $\times$ 10$^{-27}$, and 8.74 $\times$ 10$^{-27}$ erg
s$^{-1}$cm$^{-2}$Hz$^{-1}$ per pixel.

\subsection{Image Preprocessing Before Extracting the SED} 
In order to investigate the physical properties of M 101 in a 
spatially resolved manner, we plan to extract the SED pixel-by-pixel. 
As the images come from different telescopes and were taken under 
different observational conditions, some steps of image preprocessing 
(similar to \citet{zou11c}) are required: pixel scaling, masking stars, 
removing the sky background, convolution of the PSFs, and smoothing.

\subparagraph{Pixel scaling} As the images in various bands have different
spatial resolutions, we adjust these images to the same pixel scale and
direction as the BATC images (1.7{\arcsec} per pixel, north is up and east at
the left). The pixel scale of 1.7{\arcsec} corresponds to about 61 pc at the
distance of M 101.

\subparagraph{Masking stars} Foreground bright field stars are masked according
to the 2MASS Point Source Catalog \citep{skr06} and the corresponding areas are
filled with the values of their nearest pixels.

\subparagraph{Sky background subtraction} Source signals are firstly 
identified by SExtractor \citep{ber96} and masked, then the sky background 
maps are obtained by the polynomial fitting method with remaining background 
pixels.

\subparagraph{PSF convolution} The PSF for each band is calculated by the
PSFEx software\footnote{http://www.astromatic.net/software/psfex} \citep{ber11}. 
All the images are convolved to the FHWM of 6{\arcsec} in 24 $\mu$m by the 
PSFMATCH task in IRAF\footnote{http://iraf.noao.edu/}.

\subparagraph{Image smoothing} To improve the S/N for galactic outskirts, a
boxcar averaging method, as shown in \citet{kon00} and \citet{zou11c}, is
performed on all images. Here, the smoothing window size depends on the S/N 
of the BATC $i$-band image and the photometric error for each pixel in the 
specified band is estimated with the S/N.

\begin{figure} \epsscale{1.0} \plotone{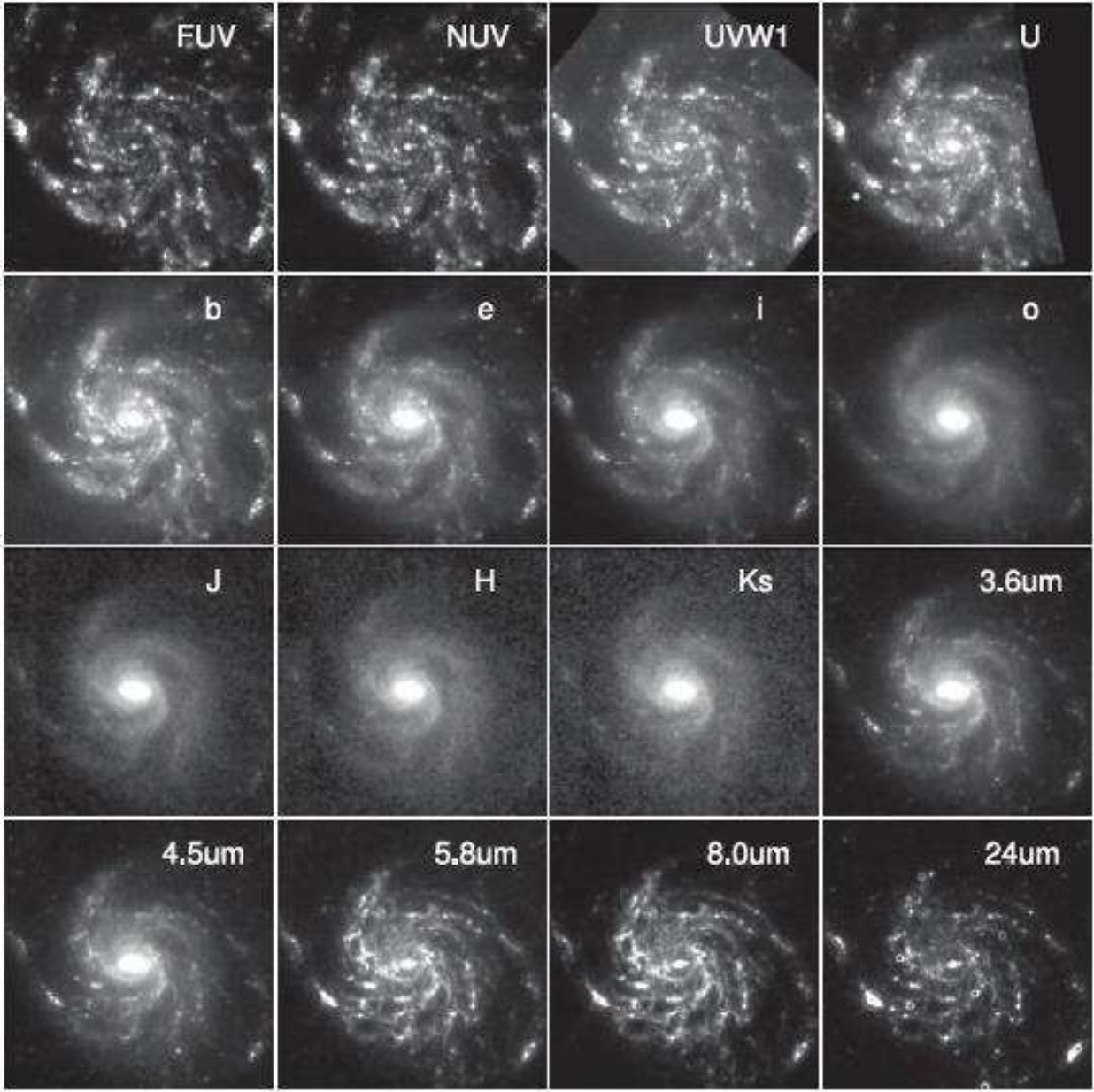} \caption{Multi-band
morphologies of M101. The image size is about 20{\arcmin} $\times$ 20{\arcmin},
corresponding to 43 $\times$ 43 kpc at the distance of 7.4 Mpc. All the images
are shown with the same spatial scale of 1.7{\arcsec}. North is up and east is
left. \label{fig1}} \end{figure}

Figure \ref{fig1} displays the processed GALEX, BATC, 2MASS, and Spitzer images
of M 101. The morphology of this galaxy belongs to the typical late-type spiral
galaxy. Star-forming spiral arms dominate the galaxy morphology at UV bands. 
Most of the UV-bright knots coincide with giant H {\sc ii} regions. The
$J$, $H$, and $K_s$ images only show features of the bulge and bright inner
disk due to short exposures. The Spitzer 3.6 and 4.5 $\mu$m trace the stellar
mass and other three bands are filled with dust radiation in addition to stellar 
light. The SED for each pixel is extracted and will be fitted with 
models as explained later. These SEDs are corrected by the foreground Galactic 
extinction, which is 0.009 mag in $E(B - V)$ \citep{sch98}.

\section{CONSTRUCTION OF SED MODELS \label{sec3}}

Stellar population synthesis fitting, including both the model prediction
and fitting procedure, has been improved significantly in the past few decades
\citep{tin72,sil98,vaz99,bru03,sal07,gro08,nol09}. This technique can be used
to effectively derive a series of physical properties of galaxies, such as
redshift, stellar mass, star formation rate (SFR), dust mass, and metallicity
\citep{dac08,zib09,ken09,zou11b,zou11c}.  In the following contents, we will
describe some aspects of the evolutionary stellar population synthesis and
create a library of SED models.

The population synthesis code of \citet{bru03}, hereafter BC03, is used to
generate evolutionary stellar population synthesis models. We adopt the IMF of
\citet{cha03} with lower and upper mass limits of 0.1 $M_\odot$ and 100
$M_\odot$. The Chabrier IMF is preferred over the Kroupa \citep{kro02} and
Salpeter \citep{sal55} IMFs because of its better agreement with the number
counts of brown dwarfs in the galactic disk \citep{mac04}. Recycling is not
considered and the ejected gas is reused in the new star formation episodes.

We choose the ``delayed-exponential'' star formation history (SFH) from the 
BC03 code. This kind of SFH ("Sandage-like" SFH; hereafter Sandage SFH) can 
describe the increased SFR and is thought to be more realistic 
\citep{gav02}. It can be expressed as 
\begin{displaymath} 
\Psi(t)=\tau^{-2}t \ \mathrm{exp}(-t/\tau)\ , 
\end{displaymath} 
where $t$ is the age, $\tau$ is the star formation timescale, and $\Psi(t)$ is
the SFR, whose maximum is located at $t=\tau$. Compared with the simple
exponentially declining SFH, the Sandage SFH is characterized by a delayed 
rise in the SFR followed by an exponential decline. The decline rate 
is determined by the value of $\tau$. \citet{mac04} showed that the two model 
grids are quite similar in the color-color space, but the Sandage SFH gives a 
younger average age ($\sim$ 0.5$-$2 Gyr) and slightly higher metallicity for 
a given color.

To denote the star formation activity, we use the $b$ parameter depicted in
\citet{sca86}. It is expressed in terms of the ratio of the present to
past-averaged star formation rate as 
\begin{displaymath} 
b = \frac{\psi(t_{0})}{\bar{\psi }} =
\frac{t_{0}\psi(t_{0})}{\int_{0}^{t_{0}}\psi(t)dt}, 
\end{displaymath} 
where $t_0$ is the current age and $\psi(t_0)$ is the SFR at $t_0$.

With the above SFH and BC03 code, we create a library of stochastic models
($\sim$10$^{5}$) following the method of \citet{kau03}. We take the age to be
uniformly distributed over the interval between 1 Myr to 13.5 Gyr. The star
formation timescale $\tau$ is distributed according to the probability density
function of $p(\gamma=1/\tau)=1-\mathrm{tanh} (8\gamma-6)$ \citep{dac08} in
order to avoid oversampling EPS models with short $\tau$, where $\gamma$ 
is uniformly sampled over the interval of 0$-$1. Metallicity is set to be uniformly
distributed in the range of 0.02$-$2 $Z_\odot$.

\section{CONSTRAINT OF THE DUST ATTENUATION} \label{sec4} 

Interstellar dust affects appearances of a galaxy in different bands. It
absorbs and scatters short-wavelength stellar light and reradiates in infrared,
making objects appear dimmer and redder. During fitting the observed SED with
models, we need to consider the dust attenuation. Since there have been
three free parameters (age, metallicity, and star formation timescale) in the
models already, we try to seek a feasible method to constrain the dust extinction 
so as to degrade the degeneracy effect among fitted parameters.

Based on the assumption of an energy balance, i.e., a fraction of UV photons
are absorbed by dust and then the energy is completely reemitted in middle IR
and FIR, the combination of the IR luminosity and rest UV luminosity might
be the most reliable way to estimate the dust attenuation \citep{bua92,meu99}.
Although almost independent on the geometry of dust and extinction law, the
ratio of the total IR (TIR) and FUV luminosity (so-called IRX in the form of
log($L_\mathrm{TIR}$/$L_\mathrm{FUV}$)) used as an estimator of the dust
attenuation, is found to depend on the age of underlying stellar population.

For star forming regions, the IRX-$A_{\mathrm{FUV}}$ relation shows an overall
consistency within $\pm$1$\sigma$ uncertainty \citep{hao11}. However, for more
evolved stellar populations, FIR emission origins from both high-energy
photons (mainly UV photons) and optical photons of intermediate-age stars,
which are both reradiated by dust. Hence, the IRX-$A_\mathrm{FUV}$ relation
might not be suitable for low-SFR populations or quiescent populations
\citep{kon04,cor08,mao12}.

With our spectral library established in the previous section, we investigate
the dependence of IRX-$A_\mathrm{FUV}$ relation on different parameters.
Each synthetic model in the library is reddened by the extinction law
of \citet{fit86}. The coefficients in the parameterized function come from
\citet{ros94}, which is adapt to M 101. This extinction law is similar to the
Galactic law except for a much weaker 2175 {\AA} bump. The extinction 
$A_\mathrm{FUV}$ is sampled in the range of $0-8$ mag. For normal star-forming 
galaxies, the typical range of $A_\mathrm{FUV}$ is $0-4$ mag.

\begin{figure}
\centering
\includegraphics[angle=0, width=0.9\textwidth]{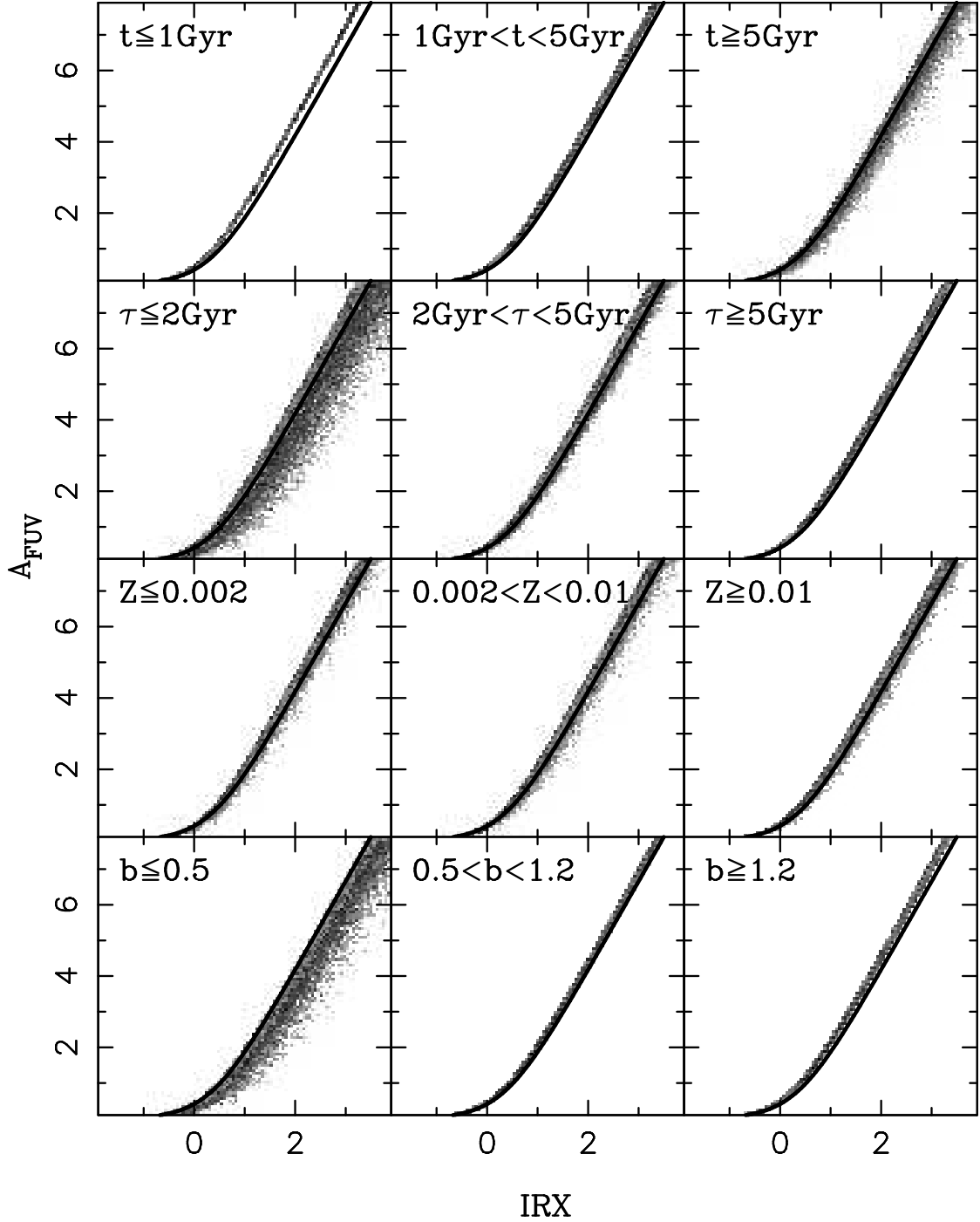}
\caption{Relationship between 
IRX and A$_{FUV}$ within different ranges of parameters. From top to bottom, 
this relation is plotted in different bins of age $t$, star formation
timescale $\tau$, metallicity $Z$, and birth rate $b$. The IRX-A$_{FUV}$
relation calibrated by \citet{hao11} is also overlapped in solid lines.
\label{fig2}} \end{figure}

Figure \ref{fig2} displays the relation between IRX and $A_\mathrm{FUV}$
within different parameter intervals. The TIR flux in IRX is calculated as the 
energy difference between the original dust-free model spectrum and corresponding 
reddened spectrum. The FUV flux is computed by convolving the reddened spectrum 
with the transmission curve of the GALEX FUV filter. The IRX-$A_\mathrm{FUV}$ 
relation calibrated by \citet{hao11} is overplotted in solid lines. From this 
figure, we can conclude that for older stellar populations ($t > 5$ Gyr), shorter 
star formation timescales ($\gamma > 0.6$ Gyr$^{-1}$), or lower birth rates 
($b < 0.5$), the IRX-$A_\mathrm{FUV}$ correlation becomes looser and the dust 
attenuation might be overestimated if just considering a simple relation 
derived from star forming regions. We also notice that the relation 
seems not change with metallicity.

To find a proper second parameter that affects the IRX-$A_\mathrm{FUV}$
relation, we introduce the formula as presented in \citet{hao11}:
\begin{equation} 
A_\mathrm{FUV}=2.5\mathrm{log}(1+\alpha \times
10^\mathrm{IRX}), \label{equ1} 
\end{equation} 
where $\alpha$ is a scale factor. The larger the value of $\alpha$ is, 
the steeper the relation curve becomes.

\begin{figure}
\centering
\includegraphics[angle=-90, width=0.9\textwidth]{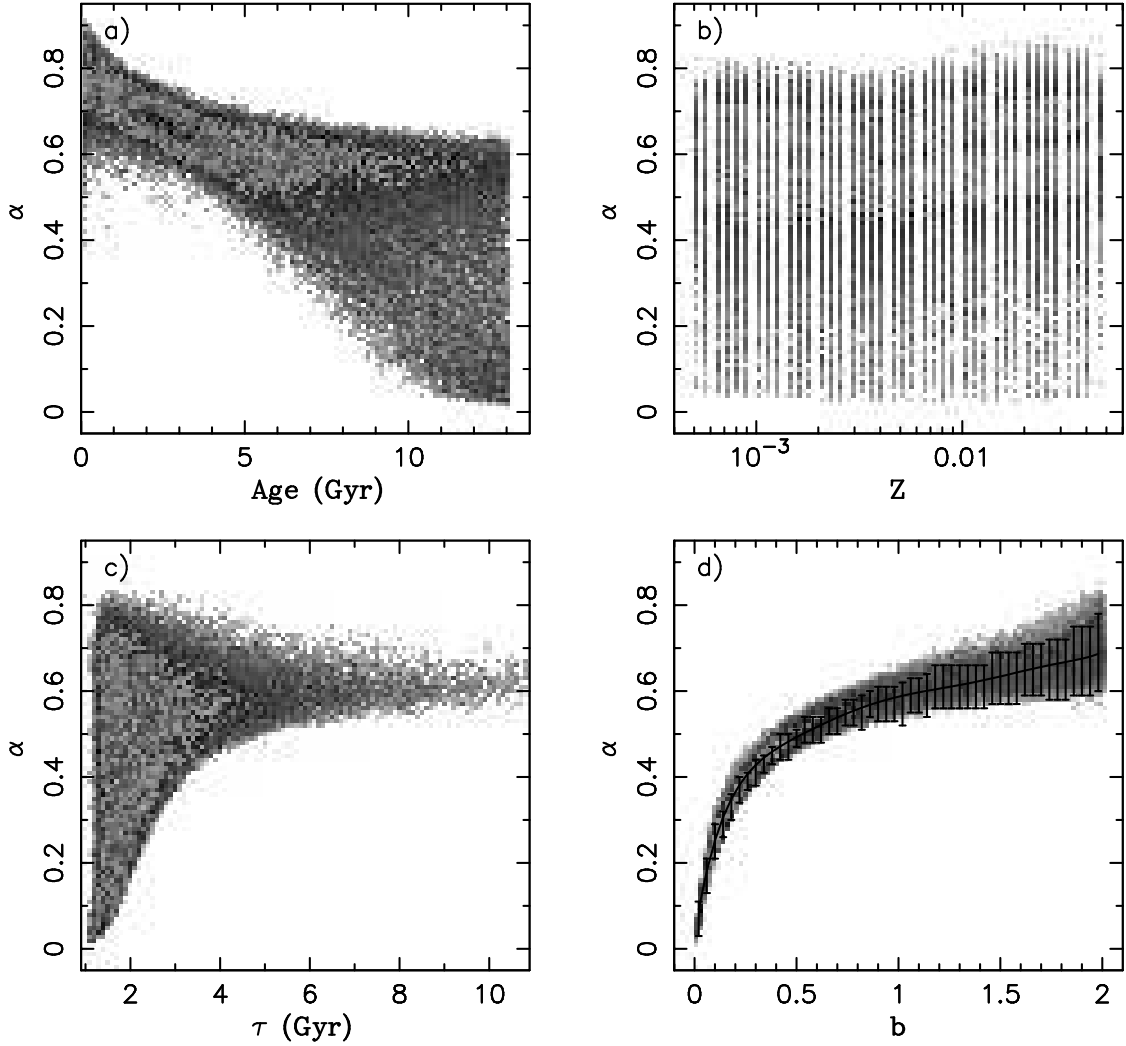}
\caption{Scale factor $\alpha$
in Equation (\ref{equ1}) as functions of age (top left), metallicity (top
right), star formation timescale (bottom left), and birth rate (bottom right).
The solid curve in the bottom right panel is the fitted polynomial as 
shown in Equation (\ref{equ2}) and points with error bars detnote means and 
stardard deviations of $\alpha$ values in given bins of $b$.  \label{fig3}} 
\end{figure}

For each model spectrum, we calculate the $\alpha$ value by using Equation
(\ref{equ1}). Figure \ref{fig3} shows the scale factor $\alpha$ as functions of
age, metallicity, star formation timescale, and birth rate. A very tight
relation between $\alpha$ and $b$ can be seen. The scale factor increases
monotonically with the birth rate, that is, the IRX-$A_\mathrm{FUV}$ relation
strongly depends on the birth rate. We fit the $\alpha$-$b$ relation with a
seven-order polynomial, shown as below: 
\begin{equation} 
\alpha=0.0377 +
2.8669b - 8.6302b^2 + 15.3222b^3 - 15.6703b^4 + 9.0869b^5 - 2.7713b^6 +
0.3451b^7. \label{equ2} 
\end{equation} 

Actually, for normal star-forming
galaxies, $\alpha$ is about 0.46 \citep{hao11}. For starburst galaxies with
larger star birth rates, $\alpha$ is about 0.6 \citep{meu99}.

\section{FITTING METHOD AND RELIABILITY TESTING} \label{sec5} 

\subsection{Chi Square Minimization} Model spectra are convolved with the
transmission curves of the UV, optical, and near infrared (3.6 and 4.5 $\mu$m)
bands to generate a set of model SEDs. A simple $\chi^{2}$ minimization of
the differences between model SEDs and observed SEDs is used to derive the
optimal values of parameters (for the convolution and minimization methods,
refer to the papers of \citet{kon00} and \citet{zou11c}).

For a given observed SED, the $\chi^{2}$ value is computed for each 
model and the optimal parameter estimates are calculated by the probability 
distribution function $p=\mathrm{exp}(-\chi^{2}/2)$ \citep{kau03,sal07}. The 
expectation value of a parameter $x$ can be expressed as \begin{displaymath} 
\bar{x} =\frac{\sum p_{i}\ x_{i}}{\sum p_{i}}, \end{displaymath} and the 
standard deviation is
\begin{displaymath} \sigma_{x} = \sqrt{\frac{\sum p_{i}\ (x_{i} -
\bar{x})^{2}}{\sum p_{i}}}, \end{displaymath} where $x$ represents $t$, $\tau$,
or $Z$.

While fitting with the models, we use the constraints of Equation (\ref{equ1})
and (\ref{equ2}) to tie the dust attenuation $A_\mathrm{FUV}$ to IRX and the
birth rate $b$ of the stellar population. Here, $b$ is to be determined. The
value of IRX is calculated according to its definition, where
\begin{displaymath} L(\mathrm{FUV}) = {\nu}L_{\nu}(\mathrm{FUV}),
\end{displaymath} and $L(\mathrm{TIR})$ is calculated with the 8 $\mu$m and 24
$\mu$m luminosities \citep{cal05}: \begin{displaymath}
\mathrm{log}L(\mathrm{TIR}) = \mathrm{log}L(24{\mu}m) + 0.908 + 0.793
\mathrm{log}\frac{L_\nu(8 {\mu}m)}{L_\nu(24 {\mu}m)}.  \end{displaymath} The 8
$\mu$m flux is dominated by both the dust radiation and stellar continuum, so
the $8 \mu$m luminosity is estimated according to the recipe of \citet{pah04}.

In the rest of this paper, we convert the extinction value of $A_\mathrm{FUV}$
to $A_V$ with the M101 extinction law in order to compare the results with 
other measurements in literatures, where $A_\mathrm{FUV} = 2.67 A_V$.

\begin{figure}
\centering
\includegraphics[angle=-0, width=0.9\textwidth]{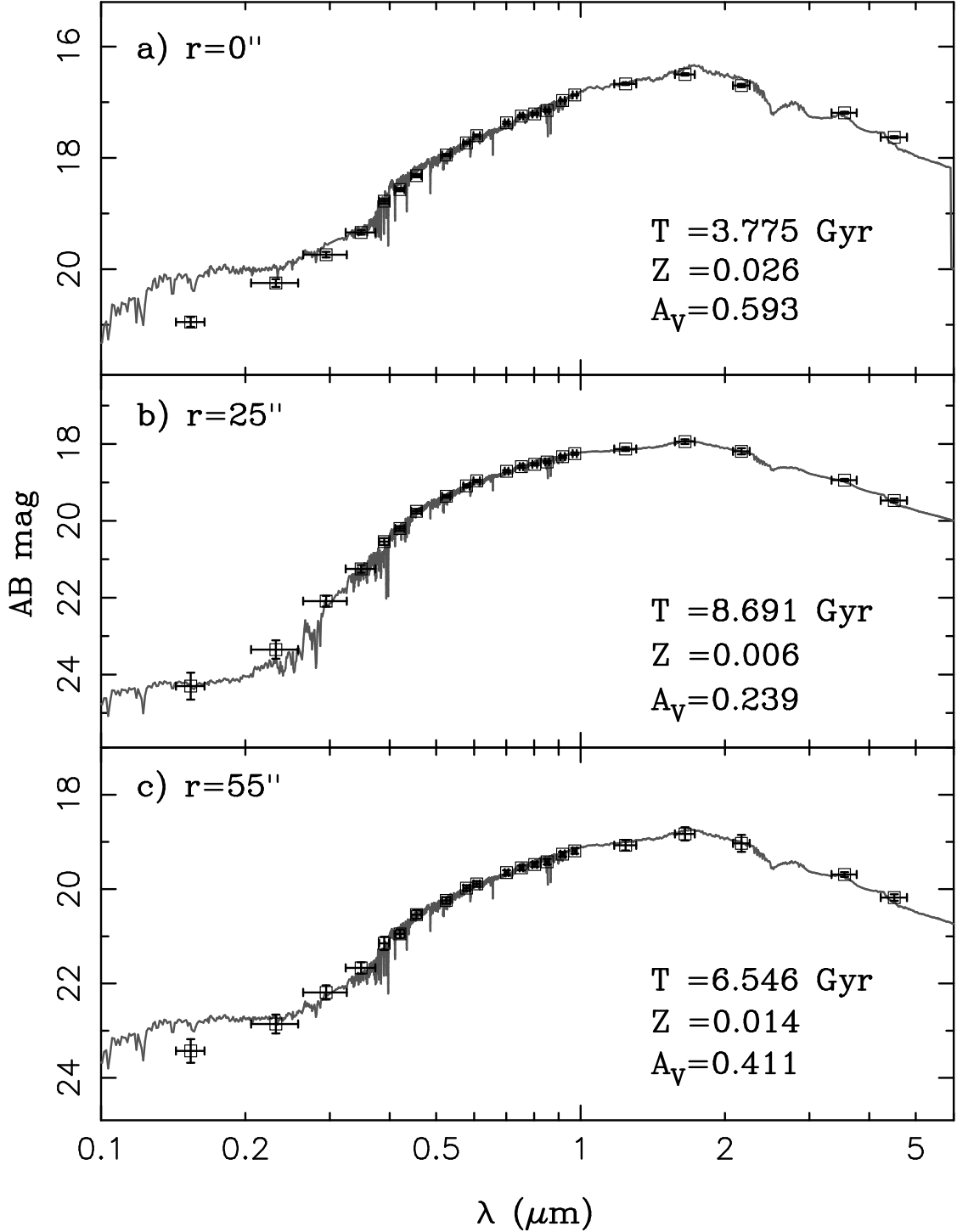}
\caption{Observed SEDs and best fitted model spectra for randomly chosen 
positions at different galactocentric radii. For each observed SED in open 
rectangles, the vertical bars indicate the photometric errors, while the 
horizontal bars show the effective bandwidths of the photometric bands. The 
model spectra are plotted in grey lines. Best fitted values of age, metallicity, 
extinction are also shown in each panel. \label{fig4}}
\end{figure}

Figure \ref{fig4} demonstrates the observed SEDs and best fitted models 
at different galactocentric radii. In this figure, we can see that the models 
fit the observed SEDs very well except for the FUV band due to lack of extreme 
horizontal-branch stars in the stellar population models \citep{maj09}. 
The upper panel of this figure presents that the center of M101 is 
younger, more metal-rich, and more dusty. The circumnuclear region as shown 
in the middle panel contains classical old bulge-like stellar populations 
and the outer region of the disk in the bottom panel is relatively younger. 

\subsection{Reliability of the Fitting Method} In order to check the
reliability of our fitting method, we create a series of artificial SED
libraries for testing. Firstly, about 1000 SEDs are generated with random
values of age, metallicity, star formation timescale, and extinction, 
which is the same process as in Section \ref{sec3}. Secondly, for a given 
S/N in the BATC $h$ band, we find the corresponding average $h$-band 
magnitude from the observed magnitude-error diagram of M 101, where the 
error $\sigma = 2.5\mathrm{log}_{10}(1 + 1/R)$, R is the S/N. All the 
artificial model SEDs are scaled to this $h$-band magnitude. Thirdly, a 
gaussian noise with the standard deviation of $\sigma$ is added to the 
scaled magnitude in each band.

\begin{figure}
\centering
\includegraphics[angle=-90, width=0.9\textwidth]{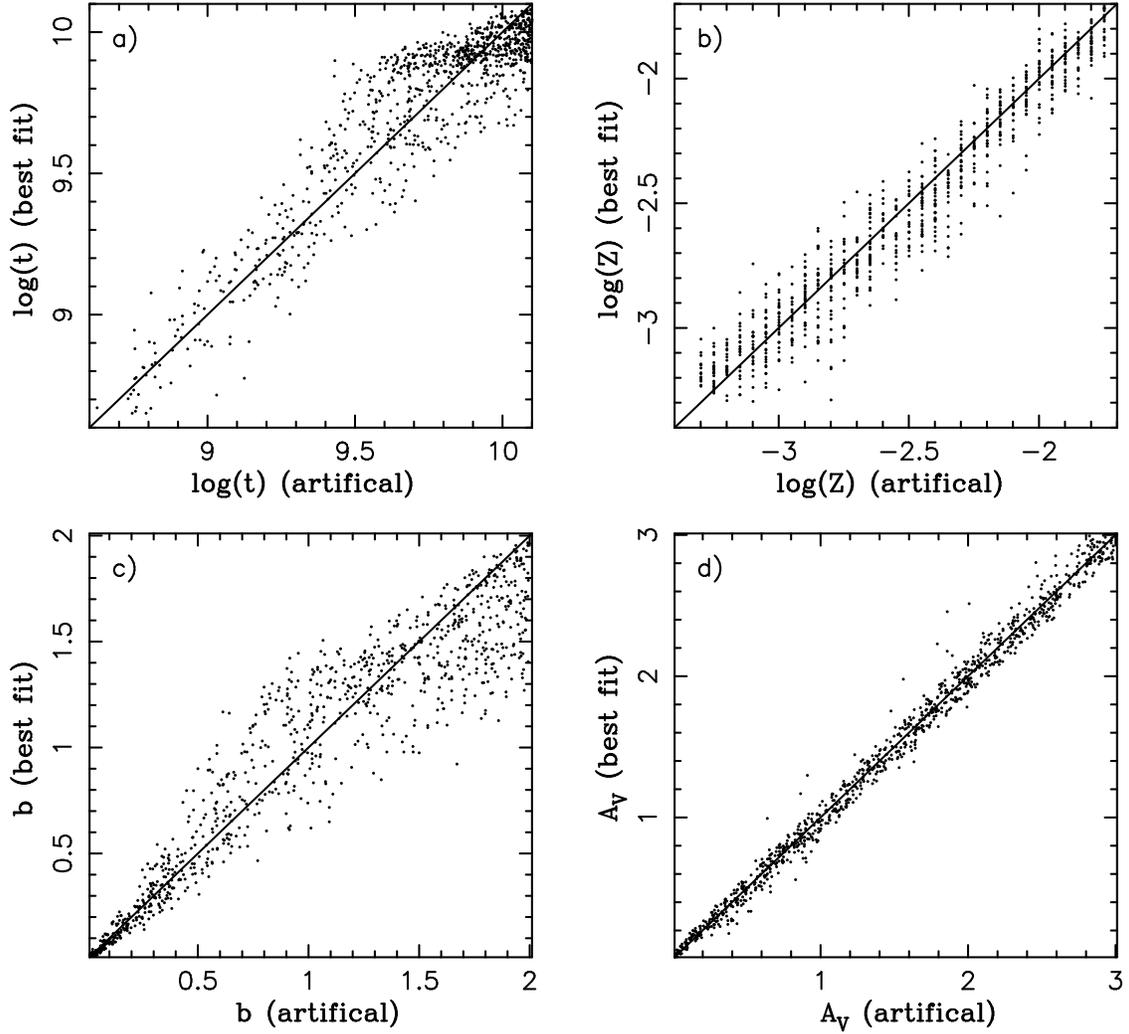}
\caption{Comparisons between
the best-fitted and input parameters after fitting the simulated SEDs (S/N of 10)
with the models. The units are yr for age, mag for $A_{V}$, and dex for
logarithmic quantities. Diagonal lines mean that the input parameters are equal
to the fitted output parameters.\label{fig5}} \end{figure}

Figure \ref{fig5} shows the capability of reproducing parameters by the
fitting method. The simulated SED library used for testing has a photometric 
S/N of 10. The resulted RMSs for age $t$ in logarithm, metallicity (log$Z$), 
birth rate $b$, and extinction $A_V$ are about 0.13 dex, 0.10 dex, 0.19, and 
0.07 mag, respectively. Another free parameter $\tau$, not drawn in this figure, 
shows a much larger scatter due to its relatively weak insensitivity to SEDs. 
Table \ref{tab2} gives parameter uncertainties estimated after fitting the 
simulated SEDs with different S/Ns.

\section{TWO DIMENSIONAL DISTRIBUTIONS OF DIFFERENT PARAMETERS} \label{sec6} 

In this section, we will show the distributions of age, metallicity, dust
extinction, and the star formation timescale of M 101 derived by the stellar
population synthesis model fitting. The deprojected radial profiles of
different parameters only relate to the regions without the effect of H {\sc
ii} regions.

\subsection{IRX and Extinction Maps} \label{sec6.1} 
\begin{figure}
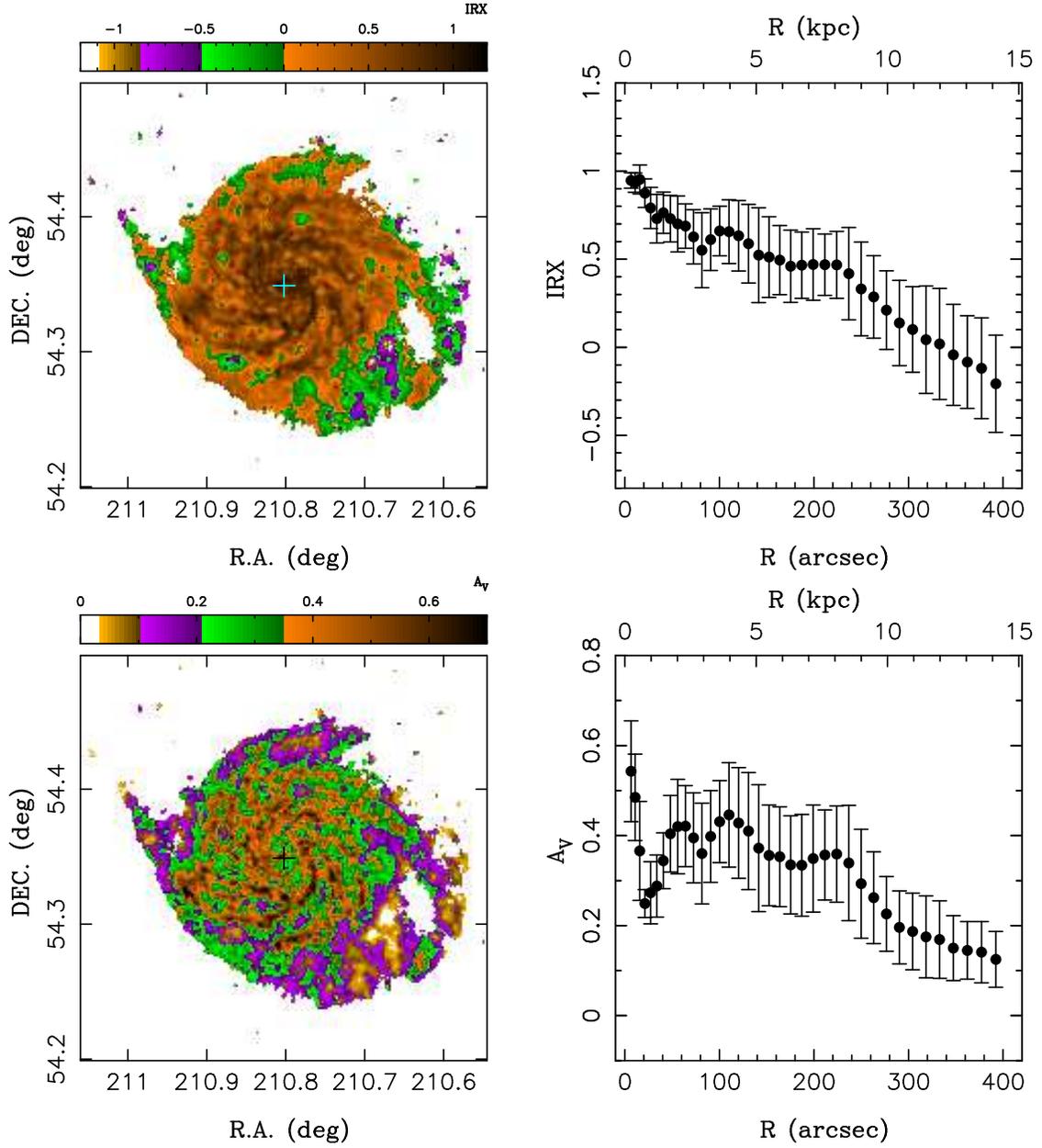

\centering
\includegraphics[angle=-90, width=0.9\textwidth]{fig6a.ps}

\includegraphics[angle=-90, width=0.9\textwidth]{fig6b.ps}
\caption{IRX and $A_V$ maps and their
corresponding radial profiles. The cross is the optical center of M 101. 
Error bars in all the radial profiles of this paper are standard deviations 
of parameter values in given annuli.
\label{fig6}} \end{figure}

Figure \ref{fig6} shows the IRX and $A_{V}$ maps and their own radial profiles. 
Both the IRX and $A_{V}$ maps present spiral arm-like structures and obvious 
radial gradients. The average extinction decreases from about 0.4 mag 
in the inner disk to about 0.05 mag at the galactic edge. Thus, the stellar 
disk of M 101 is optically thicker in the inside than in the outside. The 
central region of M 101 seems to be dusty, whose average extinction is about 
0.41 mag. The global extinction of the whole galaxy is about 0.24 mag, implying 
the optically thin interstellar medium, which is also indicated in the paper 
of \citet{boi04}. Some statistical studies show that the average extinction 
of Sc$-$Sd type galaxies is about 0.32 mag, lower than that of Sa$-$Sb type 
galaxies (0.48 mag) \citep{bos03,mun09}. In addition, from the radial profiles 
in Figure \ref{fig6}, we notice that IRX has its maximum value in the bulge (the
effective radius is about 12 arcsec \citep{fis09}) and inner disc regions,
where $A_V$ shows its minimum. It should be reasonable, because in the bulge
and inner disk, a considerable fraction of the TIR flux is contributed by old
stellar populations.

\begin{figure}
\centering
\includegraphics[angle=-90, width=0.9\textwidth]{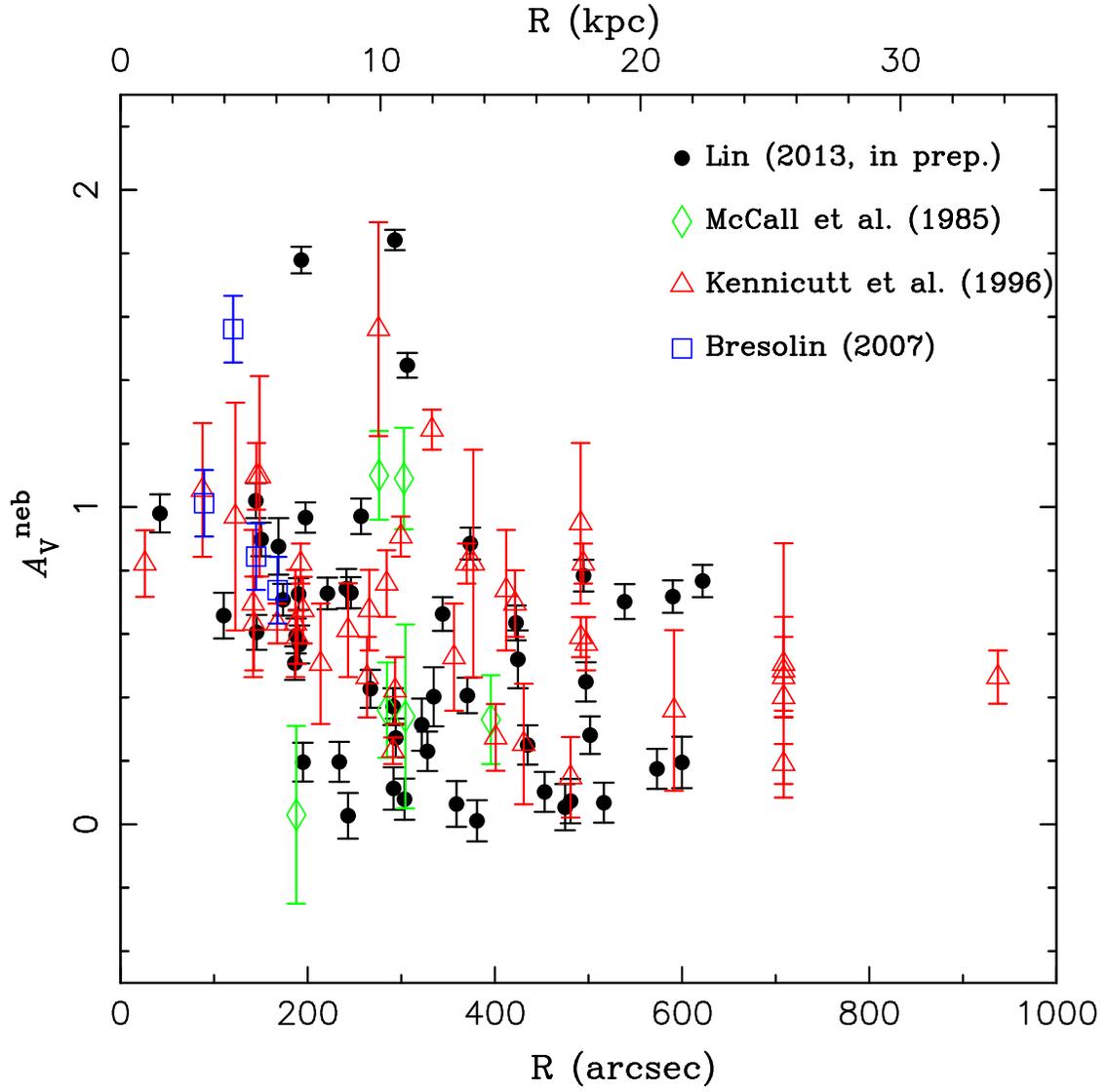}
\caption{Gas-phase attenuation
$A_{V}^\mathrm{neb}$ derived from H {\sc ii} regions. The points drawn in
different symbols come from different literatures. \label{fig7}} \end{figure}

To check the reliability of our extinction determination, we present the
extinction measurements of H {\sc ii} regions derived by the Balmer decrement 
in Figure \ref{fig7}. Here, filled circles in this figure are nebular
extinctions calculated with spectra of the Multiple Mirror Telescope (MMT).
These spectra were obtained in 2012 February under the support of the Telescope
Access Program (TAP). For detailed information about the observation and data
reduction of MMT spectra, we will publish a paper in the near future (Lin et
al. 2013; in preparation). Other extinction values in the figure come from
\citet{mcc85}, \citet{ken96}, and \citet{bre07}, where $A_\mathrm{H\alpha}$ is
converted to $A_V$ by using the M 101 extinction law. There is a rough gradient
(although somewhat diffuse) in the extinction distribution of H {\sc ii}
regions. Nebular extinctions are larger than the stellar extinctions and 
a empirical relationship was derived by \citet{cal01}: $A^\mathrm{star}_{V} = 
0.44 A^\mathrm{neb}_V$.

\begin{figure} 
\centering
\includegraphics[angle=-90, width=0.9\textwidth]{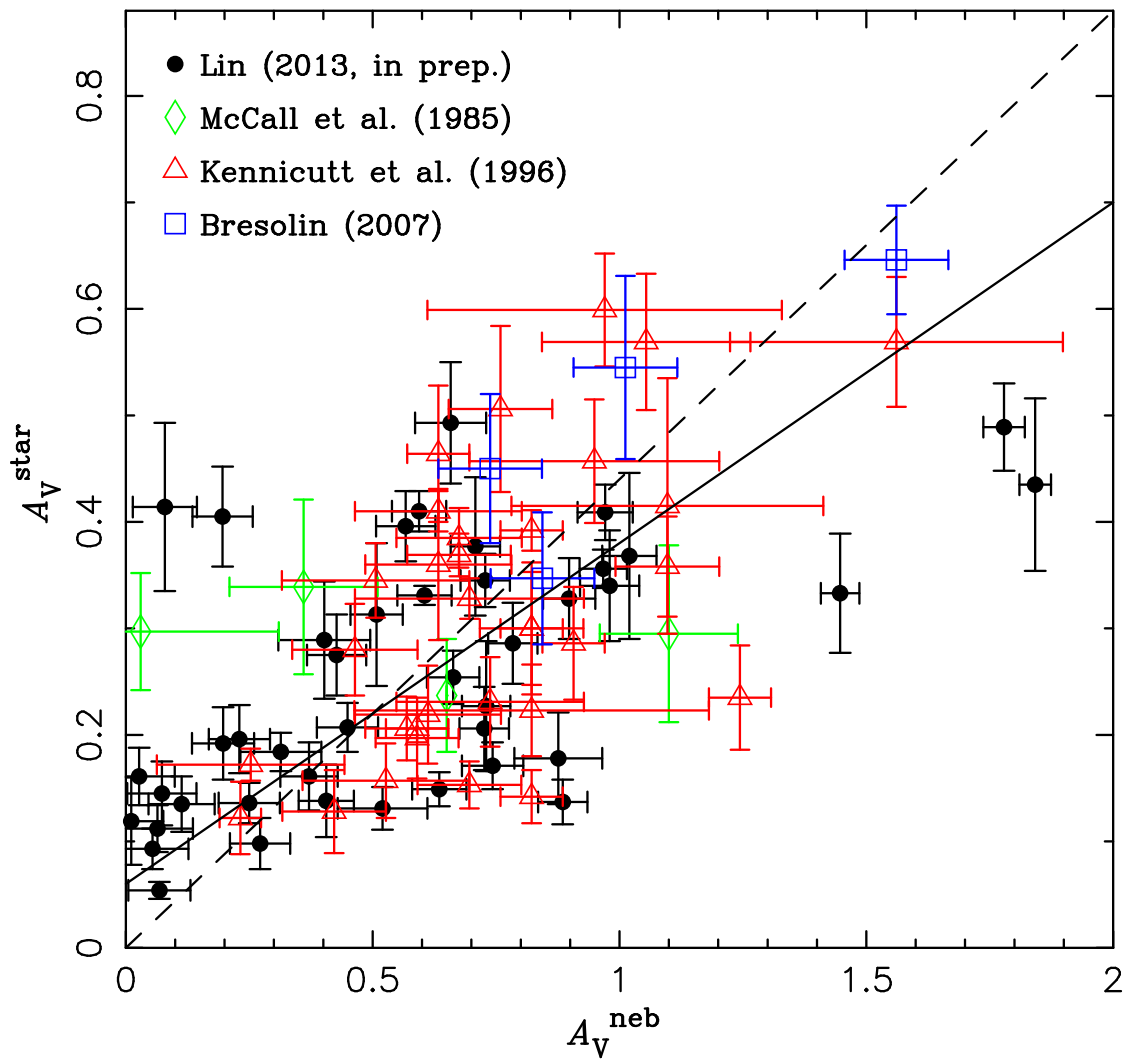}
\caption{Comparison of $A_{V}$ 
from both stellar measurements and nebular emissions in M101. The dashed line 
shows the relation from \citet{cal01}: $A^\mathrm{star}_{V}=0.44A^\mathrm{neb}_{V}$. 
The solid line displays the linear fitting to our data, which gives a flatter 
relation: $A^\mathrm{star}_{V}=0.32A^\mathrm{neb}_{V}+0.06$. \label{fig8}} 
\end{figure}

In Figure \ref{fig8}, we compare the stellar extinctions derived in 
this paper with those from nebular emission lines. The spectra of H {\sc ii} 
regions were obtained by long-slit or fiber spectrographs with apertures 
ranging from 2{\arcsec} to 10{\arcsec}. Stellar extincitons in this figure 
are determined as median values within 6{\arcsec} apertures at the same positions 
of those H {\sc ii} regions. The solid line in this figure shows a least-squares 
fit to the data points, yielding $A^\mathrm{star}_{V} = 
0.32(\pm0.01)A^\mathrm{neb}_{V} + 0.06(\pm0.01)$. The Pearson correlation 
coefficient is 0.62.  It is slightly flatter than that of \citet{cal01}, 
partly because the PSF convolution and image smoothing might 
smooth the stellar extinction.

In this paper, we simply assume a foreground screen of uniform dust and the
extinction law is same overall the galaxy. But by calculating the radiative
transfer models, \citet{pie04} discovered that there might be some different
behaviors of the attenuation function in the galaxy bulge and disk. In general,
the average slope of the extinction law increases with the opacity. A
homogeneous dust distribution would produce a little larger attenuation than a
clumpy one.

\subsection{Metallicity Map and the Radial Gradient} \label{sec6.2} 
\begin{figure}
\centering
\includegraphics[angle=-90, width=0.9\textwidth]{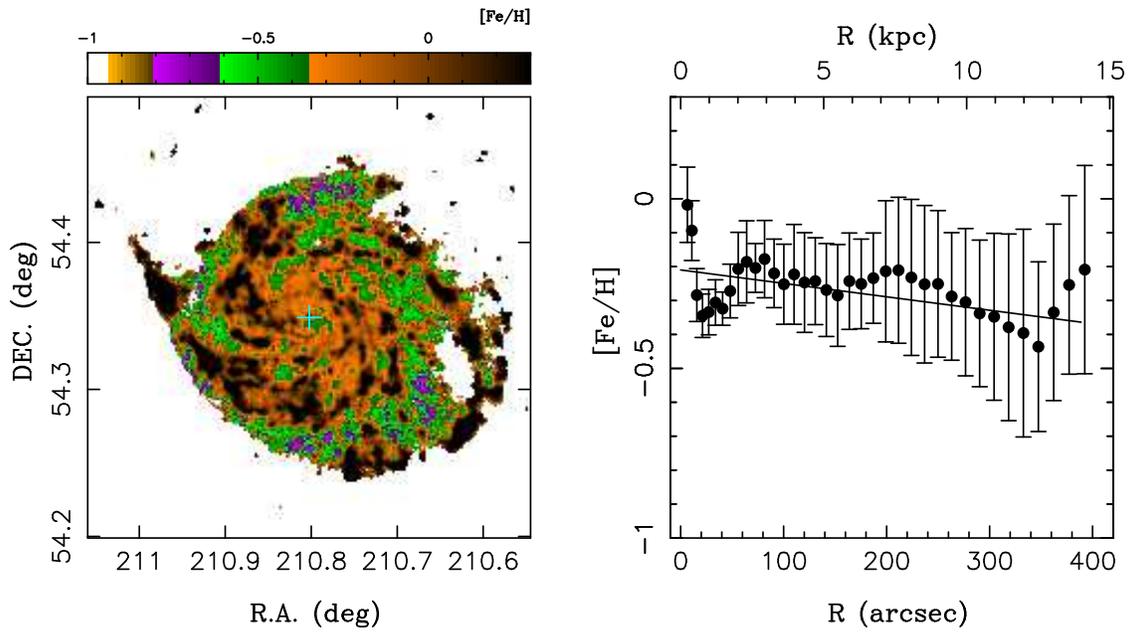}
\caption{Metallicity map in [Fe/H] and its radial profile of M 101. The cross 
is the optical center of M 101. The line in the radial profile gives the 
linearly fitted gradient of about -0.011 dex kpc$^{-1}$. \label{fig9}} 
\end{figure}

The
metallicity map in Figure \ref{fig9} shows that there are a number of knots
with high metallicities associated with the spiral arms. The dusty core is
found to be more metal-rich than the surrounding environment. The abundance in
the inter-arm regions is relatively poor. The average metallicity of the whole
galaxy is close to half of the solar abundance.

\begin{figure} 
\centering
\includegraphics[angle=-90, width=0.9\textwidth]{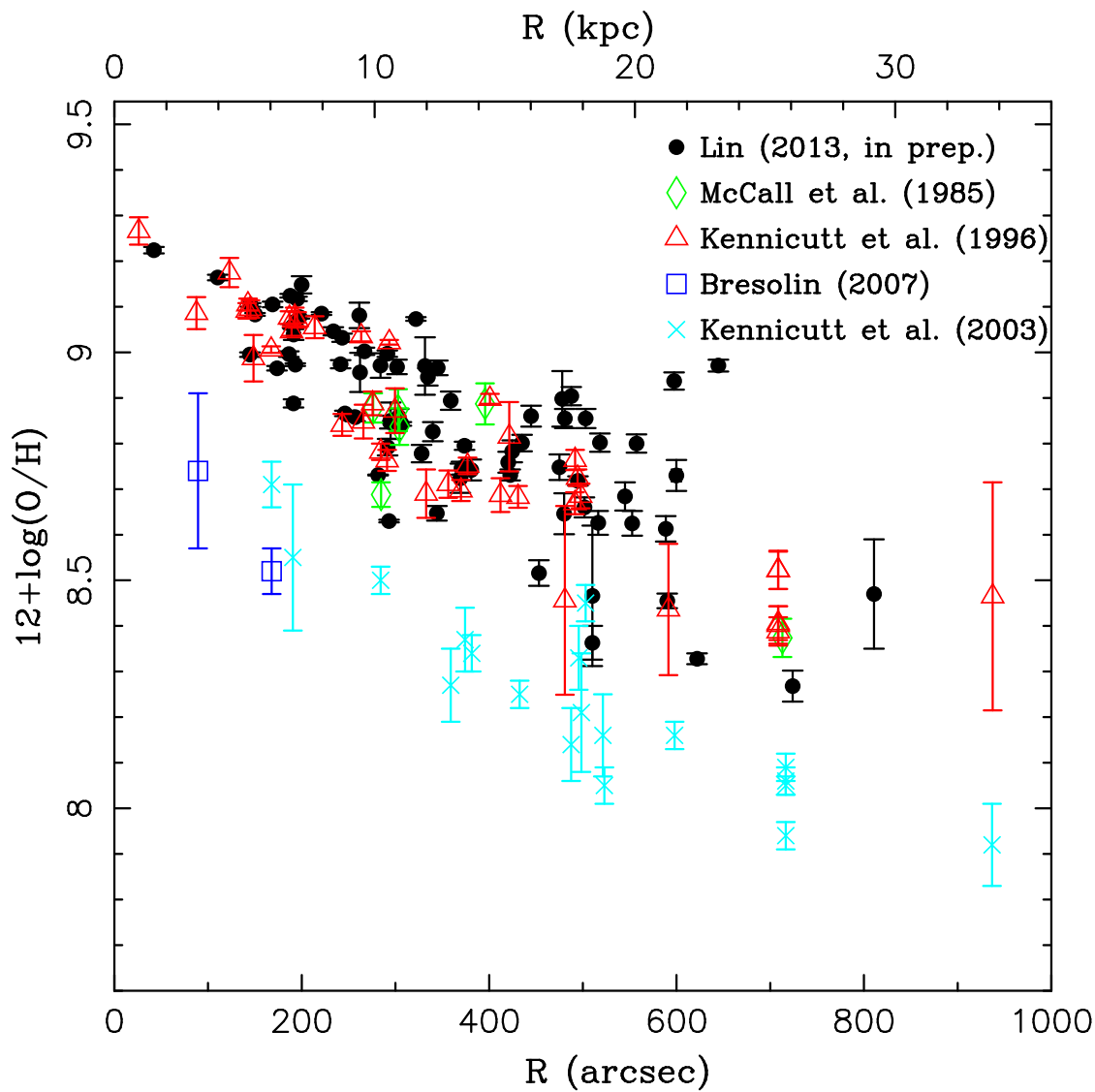}
\caption{Oxygen abundance as a
function of the deprojected galactocentric distance for individual H {\sc ii}
regions in M 101. Different symbols denote the data from different literatures.
Dark circles and orange triangles correspond to oxygen abundances computed by
the KK04 calibration \citep{kob04}. Green crosses and blue squares represent
oxygen abundances derived by the $T_{e}$ method. The KK04 calibration yields
abundances that are 0.5 dex higher than those based on the $T_{e}$ method, but
the abundance gradients are consistent based on different calibrations. \label{fig10}} 
\end{figure}

From the radial profile in Figure \ref{fig9}, there is a relatively shallow
decreasing gradient, which is about -0.011 $\pm$ 0.006 dex kpc$^{-1}$.
Gas-phase abundances of H {\sc ii} regions in M 101 have been measured by many
authors \citep{ken96,ken03,bre07}. Oxygen abundances are derived by using
different methods based on strong lines or auroral lines. We plot the collected
Oxygen abundances from the literatures along the galactocentric distance in
Figure \ref{fig10}. The metallicity measurements from the MMT spectra as
described previously are also shown in this figure. We find an average gradient
of about -0.045 dex kpc$^{-1}$, which is equal to the [Fe/H] gradient if assuming 
the same chemical composition as the Sun \citep{gre98}. In contrast, the abundance 
gradient of the circumambient stellar content is much flatter than that 
of H {\sc ii} regions.

\subsection{Age in Different Components} \label{sec6.3} 
\begin{figure}
\centering
\includegraphics[angle=-90, width=0.9\textwidth]{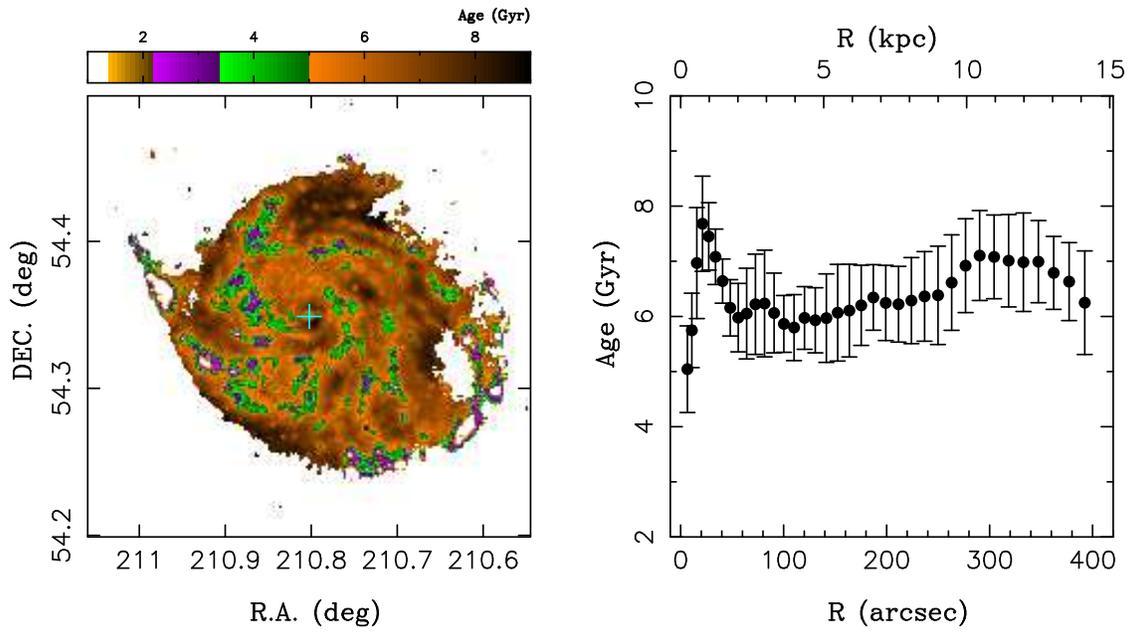}
\caption{Age map
and its radial profile. The cross is the optical center of M 101.\label{fig11}} 
\end{figure}

Figure \ref{fig11} shows
the age distribution of M 101 in two dimensions and its radial profile. The
stellar population in the central region is older, while it is younger in outer
regions. The oldest part appears in the inner disk (around the galactocentric
distance of $\sim$20 arcsec). Spiral arms, where many H {\sc ii} regions are
located, are much younger than other components. The inter-arm areas are filled
with relatively older stellar populations.

It is worthwhile to mention that the bulge has a stellar age that is younger than 
the surrounding inner disk. This kind of young bulge is also found in several 
late-type disk galaxies \citep{car07}. Similar results were reported by 
\citet{gan07} based on their IFS (Integrated Field Spectroscopy) observations 
in the central regions of late-type galaxies.

We also notice in the radial age profile that there are two distinct components
in the disk. The inner disk, ranging from 0.3 to 1.0 arcmin, is dominated by
intermediate age populations. However, the age gradient in outer disk ($R >
1.0$ arcmin) is quite flat. This kind of two parts in the disk with distinct
age gradients is also found in the studies of resolved stellar populations of
M33 and NGC 628 \citep{wil09,zou11c}.

\begin{figure} 
\centering
\includegraphics[angle=-90, width=0.9\textwidth]{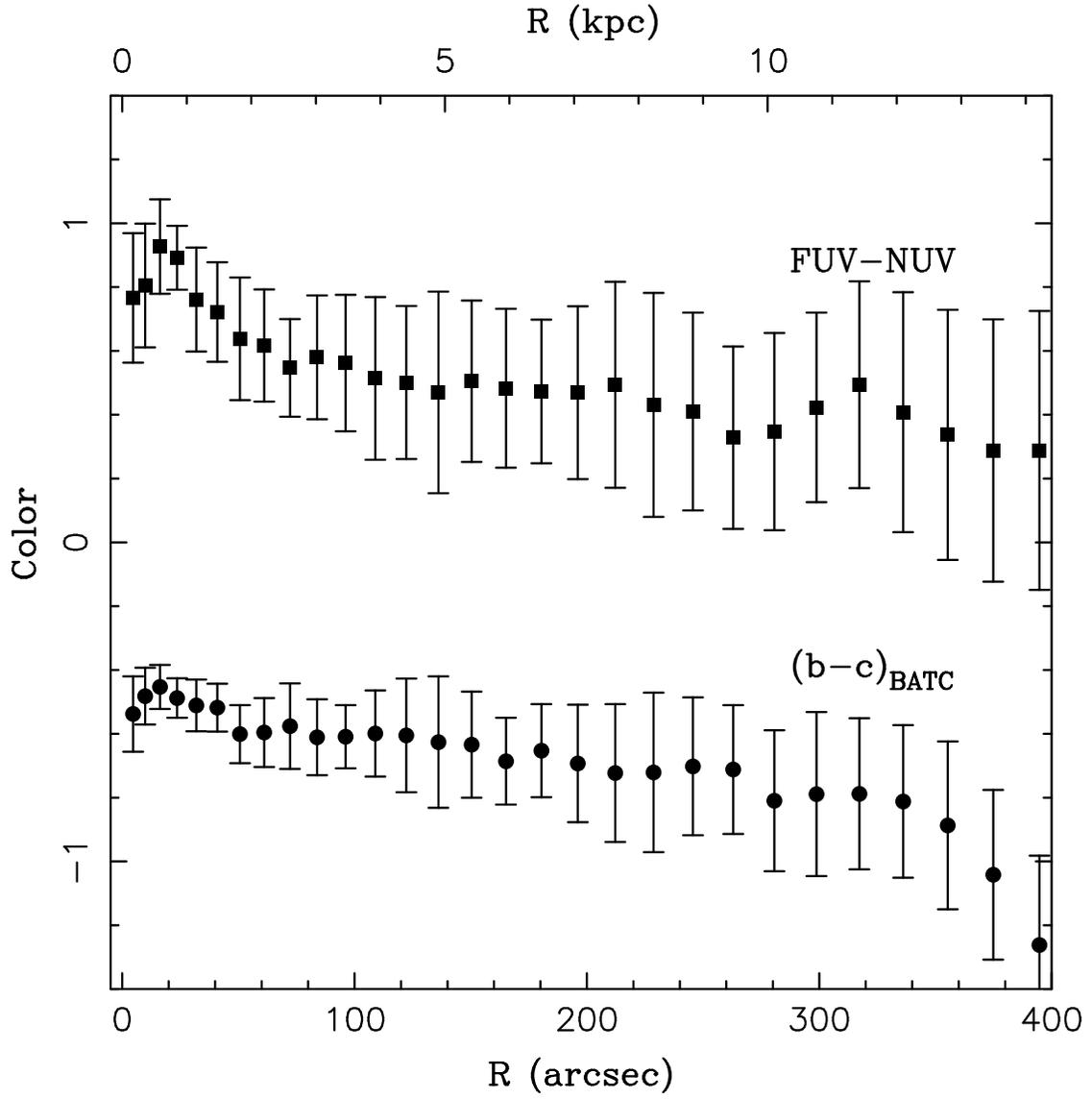}
\caption{Radial profiles of
the FUV$-$NUV and BATC $b - c$ colors. The $b - c$ color is shifted by an
arbitrary value.
\label{fig12}}
\end{figure}

Colors tracing the age are plotted in Figure \ref{fig12} in order to compare
with the age characteristics.  The FUV$-$NUV color reveals the same trend as
the age profile. As also shown in \citet{bia05}, there is a gradient of this
color in M 101, implying younger stellar populations in the outer disk regions.
Another age-sensitive index is the BATC $b - c$ color, which approximates the
index of D4000 (4000 {\AA} break). Both colors indicate that the bulge is
somewhat younger and two distinct parts in the disk presenting different color
gradients: steeper inside and flatter outside.

\subsection{Star Formation Timescale} 
\begin{figure}
\centering
\includegraphics[angle=-90, width=0.9\textwidth]{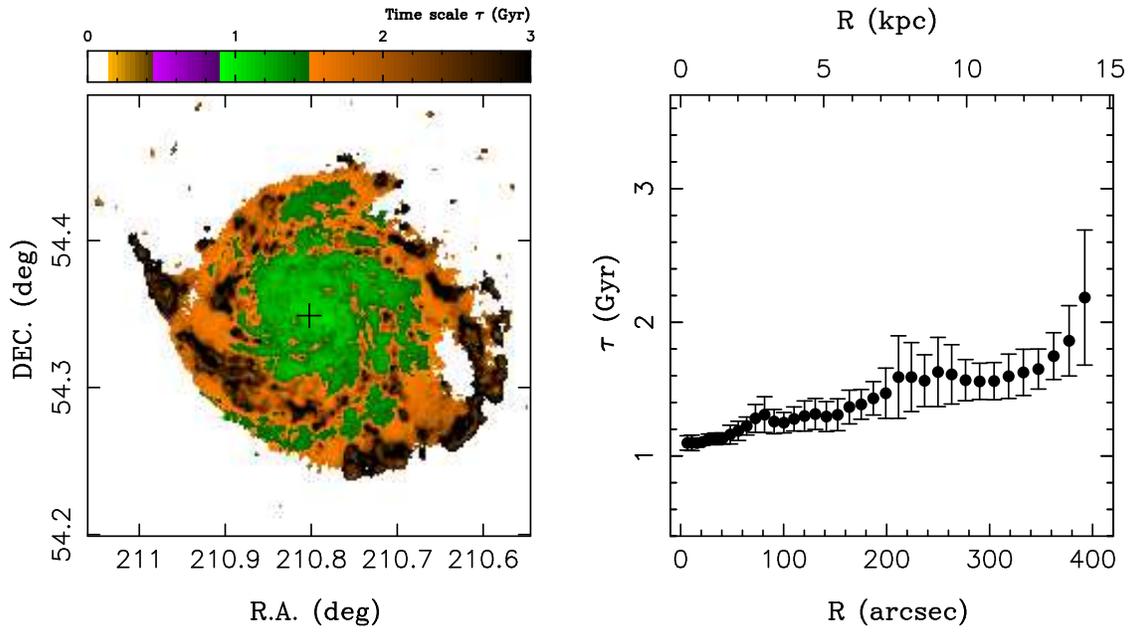}
\caption{Map of
star formation time scale and its radial profile. The cross is the optical 
center of M 101.\label{fig13}} \end{figure}

Figure \ref{fig13} shows the map of star
formation timescale and its radial profile. We can see that the timescale
increases steadily from the center to the outskirt. It is shorter in the bulge
and longer in the disk, that is, the star forming in the inner part of M 101 is
extinguished and that in the outskirt might be continuous. The average star
formation timescale of the bulge, inner disk, and outer disk is about 1.1, 1.2,
and 1.6 Gyr ($60 < R < 400$ arcsec), respectively. The average timescale of 
some H {\sc ii} regions in the outer disk may reach about 2.6 Gyr.

\section{DISCUSSION} \label{sec7} 

\subsection{Gradients} Radial surface brightness and color are the most direct
observational quantities to investigate the properties of stellar populations
and the evolution status in nearby galaxies, but it is difficult to relieve the
degeneracy effect. Using synthetic spectral models and UV-to-IR panchromatic
photometric SEDs, we have distinguished the effects of age, metallicity, and
dust extinction.

The radial age profile of M 101 as shown in Section \ref{sec6.3} presents a
comparatively younger bulge (see the discussion in next section), an older
inner region of the disk with a steeper age gradient, and a younger outer disc
region with a flatter gradient. Two distinct disc components might be caused by
differential rotations and a nonaxisymmetric self-gravitating mode of spiral
arms, which make the inner disk contract and the outer region stretch
\citep{zou11c}. Certainly, events, such as gas accretion and interacting with
another galaxy, also trigger star formation in the outer disk and make the
galaxy look younger \citep{kam92}.

The age and dust extinction maps of M101 show that the central region is 
younger, more active in star formation, and more dusty. While in the 
surrounding regions ($r$ = 20{\arcsec}), the stellar populations are 
dominated by old bulge stars and relatively dust-free. These old stellar 
populations are similar to those of typical bulges or early-type galaxies, 
which are expected to be less attenuated than star-forming galaxies \citep{cal01}.  

In our results, we find that M 101 has a radial tendency of the star formation
time scale, which is shorter in the center and longer outside. Actually, the
typical timescale in the form of the exponentially declining SFH of an
elliptical galaxy is about 1 Gyr, those of S0 and Sa$-$Sd spiral galaxies range
from 2 to 30 Gyr, and the star formation rate of an irregular galaxy is
constant \citep{bol00}. Note that for a given stellar population, the Sandage
SFH gives a much shorter timescale than the normal exponential SFH.  Inner
parts of the galaxy, such as the bulge, resemble early-type elliptical
galaxies. Most stars in these areas formed in the very early time, implying a
short timescale. The outer disk is undergoing a continuous star formation, like
late-type spiral or irregular galaxies, giving the longer timescale. Compared
with the assumption of a single star formation law (i.e., a single timescale)
overall the galaxy, a variational timescale should be more suitable for realistic
galaxies. At the same time, this kind of timescale distribution might be
important in future studies of stellar populations in different regions of
galaxies and in simulations of the galactic evolution.

There are radial metallicity gradients both in the stellar content and gas-rich
H {\sc ii} regions. The gradient of the stellar abundance is flatter and the
gas-phase gradient is much steeper. In fact, many observations of nearby spiral
galaxies show that the inner disk has higher metallicity than the outer disk
and the typical gradient of H {\sc ii} regions is about -0.05 dex kpc$^{-1}$
\citep{pil12}. The gas-phase metallicity gradient of M 101 is about -0.045 dex
kpc$^{-1}$, which is lower than the average gradient of other Sc or Scd type
galaxies, such as M33, NGC 2403, and NGC 3184 \citep{zar94}. There is some
evidence showing that interacting pairs might have systematically lower
metallicities than field galaxies \citep{kew06} and recent mergers also make
the metallicity gradient flatten. M 101 is the main body in the M 101 group. It
might be interacting with NGC 5474 and NGC 5477 in the group as suggested by
\citet{mih12a,mih12b}, who declared, with the deep neutral hydrogen observation
and deep optical images, that the group is in a dynamically active state and
encounters in the group environment are building up the outer disk of M 101. In
a whole, the existence of a metallicity gradient in M 101 supports an
inside-out disk growth scenario: the early gas infall or collapse made the
small inner region more metal-rich and the outer disk was enriched more slowly
\citep{wan11}.

From the metallicity map in Figure \ref{fig9}, we also notice that the
abundance in the inter-arm regions is poorer than spiral arms. It was expected
by some numerical models that establish density waves to trigger star
formation. \citet{roy95} suggested that some azimuthal mixing processes such as
super-shell expansion are able to wipe out the abundance inhomogeneities of the
interstellar medium in a few galactic rotation (maybe about 1$-$3 Gyr).

\subsection{Bar-induced Secular Evolution} At early times, the galactic
evolution was leaded by a combination of dissipative collapses and mergers. The
evolution timescale was short and the processes were violent
\citep{too77,san90}. In the local universe, some internal secular processes
become dominant. These secular processes tend to make late-type galaxies to be
bulgeless or have pseudobulge \citep{kor04}.

\citet{kor10} have used the Hobby-Eberly Telescope to obtain high-resolution
spectra of nuclear star clusters in M 101 and NGC 6946. They found that the
nucleus of M101 has an average rotation velocity of about 210 $\pm$ 15 km
s$^{-1}$, but the velocity dispersion is only about 25 $\pm$ 7 km s$^{-1}$.
\citet{fis10} measured the $3.6 - 8.0$ $\mu$m color as a rough estimate of the
specific star formation rate and reported that the bulge of M 101 is mildly
active. The S{\'e}rsic index of the bulge is about 1.8 \citep{fis09}, which is
less than 2. The central region of M 101 is younger than the surrounding inner
disk as shown in our age map, indicating that there is recent star forming in 
the bulge. The above characteristics, including kinematics dominated by rotation, 
small velocity dispersion, nearly exponential brightness profile ($n_b < 2$), 
and relatively active star formation, show that the bulge of M 101 is a 
so-called pseudobulge, which reserves some features of the disk.

The pseudobulge is believed to form via the internal secular evolution of the
disk \citep{kor04,fis09}. The nonaxisymmetries of the gravitational potentials
from bars, ovals, and/or spiral arms would rearrange the mass and angular
momentum and cause gas infall to build up the central mass concentration. In
normal spiral galaxies without bars and ovals, shocks actuated by the density
wave make gas in the disk lose energy and drop into the center, which gradually
forms the disk-like bulge. For barred galaxy, gravitational torques and shocks
cause the gas inflow along the bar ends to trigger star formation and make
young stellar populations and central mass convergence. For more information
about the secular evolution and pseudobulges, refer to the review of
\citet{kor04}.

The CO radio observations show that there is a bar-like morphology in the
center of M 101 \citep{ken91}. The bar feature has a length of 1{\arcmin}.5
$\pm$ 0{\arcmin}.5 with a position angle of 102$^\circ \pm 4^{\circ}$. This
feature is expected to appear at the inner Lindblad resonances (ILR). The gas
in the disk is likely to steam along the molecular bar, pile up at the ILRs,
and trigger star formation \citep{ath92,rom99}. From the maps of age,
metallicity, dust extinction, and the optical image from the Hubble Space
Telescope (HST), a bar-like knot in the center can be resolved, which is
relatively young, metal-rich, and dusty. In addition, the parameter 
discrepancies between the pseudobulge and inner parts of the disk are larger 
in the maps than in the radial profiles which might be smoothed by the bar 
feature. Thus, evolutions of the pseudobulge and disk is probably ruled 
by some secular processes of the gas-rich bar. Among late-type spirals 
including M 101, the average length of bars is small, indicating a slow 
process of gas accretion \citep{she01}. The star formation is also slower 
than early-type barred spiral galaxies \citep{mar01}.  Thus, such kind of 
secular process is less obvious in late-type spiral galaxies.

\section{SUMMARY} \label{sec8} 

Radial gradients of stellar population properties have imprints of the galactic
formation and evolution. Colors and spectral lines are usually used to
investigate the stellar population and chemical abundance of galaxies, but
there are serious degeneracy effects of age, metallicity, and dust
attenuation. Multi-band photometric data from ultraviolet, optical,
and infrared observations can help us to degrade this kind of degeneracy. By
fitting with the evolutionary stellar population synthesis models, we have
derived spatially resolved distributions of age, metallicity, dust extinction,
and star formation timescale in a nearby Scd galaxy M 101. In our fitting
method, we constrain the dust attenuation with the IRX-$A_\mathrm{FUV}$
relation, which is found to strongly correlate to a second parameter of birth
rate $b$. This parameter is related to both age and star formation history.

From the results, we find that there are clear gradients of different
parameters, supporting the so-called ``inside-out" disk growth scenario. Some
conclusions are presented as follows:

1) The stellar disk is optically thicker in the inner region than the outer
region.  The overall dust extinction of M 101 is measured to be about 0.24 mag,
which is a litter lower than galaxies of similar morphological type and much
lower than those of earlier type. The core of this galaxy is somewhat dusty.
Extinctions of H {\sc ii} regions from both our MMT observations and
literatures also show the same radial tendency. The dust attenuation 
of the stellar content is much lower than that of H {\sc ii} regions, and 
a rough linear correlation is obtained as  $A^\mathrm{star}_{V} = 0.32A^\mathrm{neb}_{V} + 0.06$.

2) The stellar metallicity has a gradient of -0.011 dec kpc$^{-1}$, lower than
that of the gas-phase abundance gradient (-0.045 dec kpc$^{-1}$) derived by the
spectra of H {\sc ii} regions from MMT observations and literatures. The
metallicity gradient of M 101 is flatter than those of Sc and Scd type
galaxies, because interactions and mergers might lower the metallicity and make
it flatten, which was suggested by \citet{kew06}.  Actually, M 101 is located in a
small group, and it might be interacting with two other members NGC 5474 and
NGC 5477 as indicated in \citet{mih12a,mih12b}.

3) Two distinct regions of the stellar disk with different age and color
gradients are discovered. The age gradient of the inner region is steeper and
that of the outer region is flatter, which might be caused by the secular
dynamical evolution of the disk.

4) We highlight that a variational star formation timescale should more exactly
describe a real galaxy than just a single timescale in the form of SFH. The
timescale of M 101 increases gradually from the galactic center to outskirts.
It should be reasonable, because the bulge in the center resembles elliptical
galaxies with small typical timescales and the outer region of the disk is
similar to late-type spirals with large timescales.

5) The bulge of M 101 is younger, more dusty, and more metal-rich than
surrounding inner disk regions, indicating that it has been forming stars
recently. Active star formation in this galaxy, together with other
observations including high rotation velocity, low velocity dispersion, and
nearly exponential brightness profile, indicates that the bulge is a so-called
pseudobulge.  From our age, metallicity, and dust maps as well as gas
distribution from radio observations, we find that there is a resolved bar in
the center. Thus, the growth of the pseudobulge and evolution of the disk are
likely to be induced by some secular processes of the bar, whose gravitational
potential causes the gas to fall through the bar ends, triggers star formation,
and makes central mass concentration.

\acknowledgments
We thank the referee for his/her thoughtful comments and insightful
suggestions that improve our paper greatly.
This work is based in part on observations made with the Spitzer Space
Telescope, which is operated by the Jet Propulsion Laboratory, California
Institute of Technology under a contract with NASA. This publication makes use
of data products from the Two Micron All Sky Survey, which is a joint project
of the University of Massachusetts and the Infrared Processing and Analysis
Center/California Institute of Technology, funded by the National Aeronautics
and Space Administration and the National Science Foundation. This research
also use data obtained through the Telescope Access Program (TAP), which is
funded by the National Astronomical Observatories of China and the Special Fund
for Astronomy from the Ministry of Finance.  This work is supported by the
National Natural Science Foundation of China (NSFC, Nos. 11073032, 11203031,
and 11225315) and Chinese Universities Scientific Fund (CUSF) and Specialized
Research Fund for the Doctoral Program of Higher Education (SRFDP, No.
20123402110037).


\clearpage

\begin{deluxetable}{lccccc} \tablewidth{0pt} \tablecaption{FIFTEEN BATC FILTERS
AND STATISTICS OF OBSERVATIONS} \tablehead{ \colhead{No.}       &
\colhead{Name\tablenotemark{a}}          &
\colhead{$\lambda_{eff}$\tablenotemark{b}}  & \colhead{Exp.
Time\tablenotemark{c}}    & \colhead{FWHM\tablenotemark{d}}        & }
\startdata 1   &   a   &   3360    &   10800   &   4.38  \\ 2   &   b   & 3890
&   14400   &   3.85  \\ 3   &   c   &   4210    &    6000   &   4.30 \\ 4   &
d   &   4550    &   18000   &   5.20  \\ 5   &   e   &   4920    & 12000   &
3.98  \\ 6   &   f   &   5270    &   10800   &   3.88  \\ 7   &   g &   5795 &
7200   &   3.83  \\ 8   &   h   &   6075    &    3600   & 5.38  \\ 9   & i   &
6660    &    7200   &   3.90  \\ 10  &   j   &   7050 &    7200   & 4.18  \\ 11
&   k   &   7490    &    7200   &   4.29  \\ 12  & m   &   8020 &    8400   &
4.85  \\ 13  &   n   &   8480    &    6000   & 3.73  \\ 14  & o   &   9190    &
10800   &   4.09  \\ 15  &   p   &   9745 &   12000   & 4.56  \\ \enddata
\tablenotetext{a}{Letters denote the filters in the BATC photometric system.}
\tablenotetext{b}{Effective wavelengths in \AA.} \tablenotetext{c}{Total
exposure times in seconds.} \tablenotetext{d}{FWHMs of the stacked images in
arcsec.} \label{tab1} \end{deluxetable}

\clearpage
\begin{deluxetable}{cccccc} \tablewidth{0pt} \tablecaption{STANDARD DEVIATIONS
OF PARAMETERS DERIVED BY FITTING SIMULATED SEDS IN DIFFERENT S/NS} 
\tablehead{\colhead{S/N} &
\colhead{$\sigma$(log $t$)} & \colhead{$\sigma$(log $Z$)}  &
\colhead{$\sigma$($\tau$)} & \colhead{$\sigma$(b)}        &
\colhead{$\sigma$(A$_{V}$)}  } \startdata 5    &   0.16  &   0.17   &   1.23  &
0.24  &  0.13   \\ 10   &   0.13  &   0.10   &   1.24  &   0.19  &  0.07   \\
20   &   0.11  &   0.06   &   1.27  &   0.16  &  0.05   \\ \enddata
\tablecomments{The units are yr for age $t$, mag for $A_V$, and dex for
logarithmic quantities. \label{tab2}} \end{deluxetable}

\clearpage

\clearpage
%
 \clearpage

 \clearpage

 \clearpage

 \clearpage

 \clearpage

 \clearpage

 \clearpage

 \clearpage

 \clearpage

\clearpage

\clearpage


\begin{thebibliography}{} 

\bibitem[Athanassoula(1992)]{ath92} Athanassoula, E.\ 1992,\mnras, 259, 345 

\bibitem[Bakos et al.(2008)]{bak08} Bakos, J., Trujillo, I.,\& Pohlen, M.\
2008, \apjl, 683, L103 

\bibitem[Bertin(2011)]{ber11} Bertin, E.\ 2011, Astronomical Data Analysis
Software and Systems XX, 442, 435

\bibitem[Bertin \& Arnouts(1996)]{ber96} Bertin, E., \& Arnouts, S.\ 1996,
\aaps, 117, 393 

\bibitem[Bianchi et al.(2005)]{bia05} Bianchi, L., Thilker, D.~A., Burgarella,
D., et al.\ 2005, \apjl, 619, L71 

\bibitem[Boissier et al.(2004)]{boi04} Boissier, S., Boselli, A., Buat, V.,
Donas, J., \& Milliard, B.\ 2004, \aap, 424, 465 

\bibitem[Bolzonella et al.(2000)]{bol00} Bolzonella, M., Miralles, J.-M., \&
Pell{\'o}, R.\ 2000, \aap, 363, 476 

\bibitem[Boselli et al.(2003)]{bos03} Boselli, A., Gavazzi, G., \& Sanvito, G.\
2003, \aap, 402, 37

\bibitem[Bosma et al.(1981)]{bos81} Bosma, A., Goss, W.~M., \& Allen, R.~J.\
1981, \aap, 93, 106 

\bibitem[Bresolin(2007)]{bre07} Bresolin, F.\ 2007, \apj, 656, 186 

\bibitem[Bruzual \& Charlot(2003)]{bru03} Bruzual, G., \& Charlot, S.\ 2003,
\mnras, 344, 1000 

\bibitem[Buat(1992)]{bua92} Buat, V.\ 1992, \aap, 264, 444 

\bibitem[Calzetti(2001)]{cal01} Calzetti, D.\ 2001, \pasp, 113, 1449 

\bibitem[Calzetti et al.(2005)]{cal05} Calzetti, D., Kennicutt, R.~C., Jr.,
Bianchi, L., et al.\ 2005, \apj, 633, 871 

\bibitem[Cameron et al.(2009)]{cam09} Cameron, E., Driver, S.~P., Graham,
A.~W., \& Liske, J.\ 2009, \apj, 699, 105

\bibitem[Carollo et al.(2007)]{car07} Carollo, C.~M., Scarlata, C., Stiavelli,
M., Wyse, R.~F.~G., \& Mayer, L.\ 2007, \apj, 658, 960

\bibitem[Chabrier(2003)]{cha03} Chabrier, G.\ 2003, \pasp, 115, 763

\bibitem[Cortese et al.(2008)]{cor08} Cortese, L., Boselli, A., Franzetti, P.,
et al.\ 2008, \mnras, 386, 1157 

\bibitem[da Cunha et al.(2008)]{dac08} da Cunha, E., Charlot, S., \& Elbaz, D.\
2008, \mnras, 388, 1595 

\bibitem[Daddi et al.(2005)]{dad05} Daddi, E., Renzini, A., Pirzkal, N., et
al.\ 2005, \apj, 626, 680 

\bibitem[Fan et al.(1996)]{fan96} Fan, X., Burstein, D., Chen, J.-S., et al.\
1996, \aj, 112, 628 

\bibitem[Fisher \& Drory(2010)]{fis10} Fisher, D.~B., \& Drory, N.\ 2010, \apj,
716, 942 \bibitem[Fisher et al.(2009)]{fis09} Fisher, D.~B., Drory, N., \&
Fabricius, M.~H.\ 2009, \apj, 697, 630 

\bibitem[Fitzpatrick(1986)]{fit86} Fitzpatrick, E.~L.\ 1986, \aj, 92, 1068
 
\bibitem[Ganda et al.(2007)]{gan07} Ganda, K., Peletier, R.~F., McDermid,
R.~M., et al.\ 2007, \mnras, 380, 506

\bibitem[Gavazzi et al.(2002)]{gav02} Gavazzi, G., Bonfanti, C., Sanvito, G.,
Boselli, A., \& Scodeggio, M.\ 2002, \apj, 576, 135 

\bibitem[Grevesse \& Sauval(1998)]{gre98} Grevesse, N., \& Sauval, A.~J.\ 1998, \ssr, 85, 161

\bibitem[Groves et al.(2008)]{gro08} Groves, B., Dopita, M.~A., Sutherland,
R.~S., et al.\ 2008, \apjs, 176, 438 

\bibitem[Hao et al.(2011)]{hao11} Hao, C.-N., Kennicutt, R.~C., Johnson, B.~D.,
et al.\ 2011, \apj, 741, 124 

\bibitem[Jarrett et al.(2003)]{jar03} Jarrett, T.~H., Chester, T., Cutri, R.,
Schneider, S.~E., \& Huchra, J.~P.\ 2003, \aj, 125, 525 

\bibitem[Kamphuis \& Briggs(1992)]{kam92} Kamphuis, J., \& Briggs, F.\ 1992,
\aap, 253, 335 

\bibitem[Kauffmann et al.(2003)]{kau03} Kauffmann, G., Heckman, T.~M., White,
S.~D.~M., et al.\ 2003, \mnras, 341, 33 

\bibitem[Kenney et al.(1991)]{ken91} Kenney, J.~D.~P., Scoville, N.~Z., \&
Wilson, C.~D.\ 1991, \apj, 366, 432 

\bibitem[Kennicutt et al.(2003)]{ken03} Kennicutt, R.~C., Jr., Bresolin, F., \&
Garnett, D.~R.\ 2003, \apj, 591, 801 

\bibitem[Kennicutt et al.(2007)]{ken07} Kennicutt, R.~C., Jr., Calzetti, D.,
Walter, F., et al.\ 2007, \apj, 671, 333 

\bibitem[Kennicutt \& Garnett(1996)]{ken96} Kennicutt, R.~C., Jr., \& Garnett,
D.~R.\ 1996, \apj, 456, 504 

\bibitem[Kennicutt et al.(2009)]{ken09} Kennicutt, R.~C., Jr., Hao, C.-N.,
Calzetti, D., et al.\ 2009, \apj, 703, 1672 

\bibitem[Kewley et al.(2006)]{kew06} Kewley, L.~J., Geller, M.~J., \& Barton,
E.~J.\ 2006, \aj, 131, 2004 

\bibitem[Kobulnicky \& Kewley(2004)]{kob04} Kobulnicky, H.~A., \& Kewley,
L.~J.\ 2004, \apj, 617, 240 

\bibitem[Kong et al.(2000)]{kon00} Kong, X., Zhou, X., Chen, J., et al.\ 2000,
\aj, 119, 2745 

\bibitem[Kong et al.(2004)]{kon04} Kong, X., Charlot, S., Brinchmann, J., \&
Fall, S.~M.\ 2004, \mnras, 349, 769 

\bibitem[Kormendy et al.(2010)]{kor10} Kormendy, J., Drory, N., Bender, R., \&
Cornell, M.~E.\ 2010, \apj, 723, 54 

\bibitem[Kormendy \& Kennicutt(2004)]{kor04} Kormendy, J., \& Kennicutt, R.~C.,
Jr.\ 2004, \araa, 42, 603 

\bibitem[Kroupa(2002)]{kro02} Kroupa, P.\ 2002, Science, 295, 82

\bibitem[Kuntz et al.(2008)]{kun08} Kuntz, K.~D., Harrus, I., McGlynn, T.~A.,
Mushotzky, R.~F., \& Snowden, S.~L.\ 2008, \pasp, 120, 740 

\bibitem[Lee et al.(2008)]{lee08} Lee, J.~C., Kennicutt, R.~C., Engelbracht,
C.~W., et al.\ 2008, Formation and Evolution of Galaxy Disks, 396, 151 

\bibitem[Li et al.(2004)]{li04} Li, J.-L., Zhou, X., Ma, J., \& Chen, J.-S.\
2004, \cjaa, 4, 143 

\bibitem[MacArthur et al.(2004)]{mac04} MacArthur, L.~A., Courteau, S., Bell,
E., \& Holtzman, J.~A.\ 2004, \apjs, 152, 175 

\bibitem[MacArthur et al.(2009)]{mac09} MacArthur, L.~A., Gonz{\'a}lez, J.~J.,
\& Courteau, S.\ 2009, \mnras, 395, 28 

\bibitem[Ma et al.(2009)]{maj09} Ma, J., de Grijs, R., Fan, Z., et al.\ 2009, Research in Astronomy and Astrophysics, 9, 641 

\bibitem[Martin \& Kennicutt(2001)]{mar01} Martin, C.~L., \& Kennicutt, R.~C.,
Jr.\ 2001, \apj, 555, 301 

\bibitem[Martin et al.(2005)]{mar05} Martin, D.~C., Fanson, J., Schiminovich,
D., et al.\ 2005, \apjl, 619, L1 

\bibitem[McCall et al.(1985)]{mcc85} McCall, M.~L., Rybski, P.~M., \& Shields,
G.~A.\ 1985, \apjs, 57, 1 

\bibitem[Meurer et al.(1999)]{meu99} Meurer, G.~R., Heckman, T.~M., \&
Calzetti, D.\ 1999, \apj, 521, 64 

\bibitem[Mihos et al.(2012a)]{mih12a} Mihos, C., Harding, P., Spengler, C.,
Rudick, C., \& Feldmeier, J.\ 2012a, arXiv:1211.3095 

\bibitem[Mihos et al.(2012b)]{mih12b} Mihos, C., Keating, K.,
Holley-Bockelmann, K., Pisano, D.~J., \& Kassim, N.\ 2012b, arXiv:1210.8333 

\bibitem[Mao et al.(2012)]{mao12} Mao, Y.-W., Kennicutt, R.~C., Jr., Hao,
C.-N., Kong, X., \& Zhou, X.\ 2012, \apj, 757, 52 

\bibitem[Mu{\~n}oz-Mateos et al.(2009)]{mun09} Mu{\~n}oz-Mateos, J.~C., Gil de
Paz, A., Boissier, S., et al.\ 2009, \apj, 701, 1965

\bibitem[Noll et al.(2009)]{nol09} Noll, S., Burgarella, D., Giovannoli, E., et
al.\ 2009, \aap, 507, 1793 

\bibitem[Pahre et al.(2004)]{pah04} Pahre, M.~A., Ashby, M.~L.~N., Fazio,
G.~G., \& Willner, S.~P.\ 2004, \apjs, 154, 229

\bibitem[Peletier et al.(1990)]{pel90} Peletier, R.~F., Davies, R.~L.,
Illingworth, G.~D., Davis, L.~E., \& Cawson, M.\ 1990, \aj, 100, 1091

\bibitem[Pierini et al.(2004)]{pie04} Pierini, D., Gordon, K.~D., Witt, A.~N.,
\& Madsen, G.~J.\ 2004, \apj, 617, 1022 

\bibitem[Pilkington et al.(2012)]{pil12} Pilkington, K., Few, C.~G., Gibson,
B.~K., et al.\ 2012, \aap, 540, A56 

\bibitem[Ro{\v s}kar et al.(2008)]{ros08} Ro{\v s}kar, R., Debattista, V.~P.,
Stinson, G.~S., et al.\ 2008, \apjl, 675, L65

\bibitem[Roming et al.(1999)]{rom99} Roming, P.~W.~A., Moody, J.~W., \& Hintz,
E.~G.\ 1999, \aj, 117, 1733 

\bibitem[Rosa \& Benvenuti(1994)]{ros94} Rosa, M.~R., \& Benvenuti, P.\ 1994,
\aap, 291, 1 

\bibitem[Roy \& Kunth(1995)]{roy95} Roy, J.-R., \& Kunth, D.\ 1995, \aap, 294,
432 

\bibitem[Salim et al.(2007)]{sal07} Salim, S., Rich, R.~M., Charlot, S., et
al.\ 2007, \apjs, 173, 267 

\bibitem[Salpeter(1955)]{sal55} Salpeter, E.~E.\ 1955, \apj, 121, 161

\bibitem[S{\'a}nchez-Bl{\'a}zquez et al.(2009)]{san09}
S{\'a}nchez-Bl{\'a}zquez, P., Courty, S., Gibson, B.~K., \& Brook, C.~B.\ 2009,
\mnras, 398, 591 

\bibitem[Sandage(1990)]{san90} Sandage, A.\ 1990, \jrasc, 84, 70 

\bibitem[Scalo \& Struck-Marcell(1986)]{sca86} Scalo, J.~M., \& Struck-Marcell,
C.\ 1986, \apj, 301, 77 

\bibitem[Schlegel et al.(1998)]{sch98} Schlegel, D.~J., 
Finkbeiner, D.~P., \& Davis, M.\ 1998, \apj, 500, 525 

\bibitem[Sheth(2001)]{she01} Sheth, K.\ 2001, The Central Kiloparsec of
Starbursts and AGN: The La Palma Connection, 249, 605 

\bibitem[Silva et al.(1998)]{sil98} Silva, L., Granato, G.~L., Bressan, A., \&
Danese, L.\ 1998, \apj, 509, 103 

\bibitem[Skrutskie et al.(2006)]{skr06} Skrutskie, M.~F., Cutri, R.~M.,
Stiening, R., et al.\ 2006, \aj, 131, 1163 

\bibitem[Tamura \& Ohta(2003)]{tam03} Tamura, N., \& Ohta, K.\ 2003, \aj, 126,
596 

\bibitem[Tinsley(1972)]{tin72} Tinsley, B.~M.\ 1972, \aap, 20, 383 

\bibitem[Toomre(1977)]{too77} Toomre, A.\ 1977, Evolution of Galaxies and
Stellar Populations, 401 

\bibitem[Trujillo et al.(2006)]{tru06} Trujillo, I., F{\"o}rster Schreiber,
N.~M., Rudnick, G., et al.\ 2006, \apj, 650, 18

\bibitem[van Zee et al.(1998)]{van98} van Zee, L., Salzer, J.~J., Haynes,
M.~P., O'Donoghue, A.~A., \& Balonek, T.~J.\ 1998, \aj, 116, 2805

\bibitem[Vazdekis(1999)]{vaz99} Vazdekis, A.\ 1999, \apj, 513, 224

\bibitem[Wang et al.(2011)]{wan11} Wang, J., Kauffmann, G., Overzier, R., et
al.\ 2011, \mnras, 412, 1081 

\bibitem[Williams et al.(2009)]{wil09} Williams, B.~F., Dalcanton, J.~J.,
Dolphin, A.~E., Holtzman, J., \& Sarajedini, A.\ 2009, \apjl, 695, L15 

\bibitem[Worthey(1994)]{wor94} Worthey, G.\ 1994, \apjs, 95, 107 

\bibitem[Wyse et al.(2006)]{wys06} Wyse, R.~F.~G., Gilmore, G., Norris, J.~E.,
et al.\ 2006, \apjl, 639, L13 

\bibitem[Zaritsky et al.(1994)]{zar94} Zaritsky, D., Kennicutt, R.~C., Jr., \&
Huchra, J.~P.\ 1994, \apj, 420, 87

\bibitem[Zibetti et al.(2009)]{zib09} Zibetti, S., Charlot,S., \& Rix, H.-W.\
2009, \mnras, 400, 1181 

\bibitem[Zou(2011a)]{zou11a} Zou, H.\ 2011a, \pasp, 123, 1135 

\bibitem[Zou et al.(2011b)]{zou11b} Zou, H., Yang, Y.-B., Zhang, T.-M., et al.\
2011b, Research in Astronomy and Astrophysics, 11, 1093

\bibitem[Zou et al.(2011c)]{zou11c} Zou, H., Zhang, W., Yang, Y., et al.\
2011c, \aj, 142, 16

\end{thebibliography}
\end{document}